\documentclass[11pt]{article}
\usepackage{amsmath}
\usepackage{amsthm}
\usepackage[pdftex]{graphicx}
\usepackage{psfrag,epsf}
\usepackage{enumerate}
\usepackage{natbib}
\usepackage{subcaption}
\usepackage[colorlinks,citecolor=blue,urlcolor=blue]{hyperref}
\usepackage[boxruled,scleft]{algorithm2e}
\usepackage[top=1.2in, bottom=1.2in, left=1.2in, right=1.2in]{geometry}
\allowdisplaybreaks

\RestyleAlgo{boxruled}

\numberwithin{equation}{section}

\newtheorem{thm}{Theorem}

\begin{document}

\title{ \textbf{Null Models and Community Detection in Multi-Layer Networks}\footnote{This work was supported in part by  National Science Foundation grants DMS-1830547, DMS-2015561 and CCF-1934986. We also thank two anonymous reviewers for their extensive comments which have immensely helped us improve the paper.}
}

\vspace{0.3in}

\author{\textbf{Subhadeep Paul}, \textit{The Ohio State University}  \\
\textbf{Yuguo Chen}, \textit{University of Illinois at Urbana-Champaign}}
\date{}
\maketitle







\begin{abstract}
Multi-layer or multiplex networks represent relational data on a set of entities (nodes) with multiple types of relations (edges) among them where each type of relation is represented as a network layer. 
A large group of popular community detection methods in networks are based on optimizing a quality function known as the modularity score, which is a measure of the extent of presence of module or community structure in networks compared to a suitable null model. Here we introduce several multi-layer network modularity and model likelihood quality function measures using different null models of the multi-layer network, motivated by empirical observations in networks from a diverse field of applications. In particular, we define multi-layer variants of the Chung-Lu expected degree model as null models that differ in their modeling of the multi-layer degrees. We propose simple estimators for the models and prove their consistency properties. A hypothesis testing procedure is also proposed for selecting an appropriate null model for data.
These null models are used to define modularity measures as well as model likelihood based quality functions. The proposed measures are then optimized to detect the optimal community assignment of nodes \footnote{Code available at: \url{https://u.osu.edu/subhadeep/codes/}}.  We compare the effectiveness of the measures in community detection in simulated networks and then apply them to four real multi-layer networks.
\end{abstract}

\noindent%

\section{Introduction}

The complex networks encountered in biology, social sciences, economics and machine learning are often ``multi-layered" or ``multi-relational" in the sense that they consist of multiple types of edges/relations among a group of entities. Each of those different types of relation can be viewed as creating its own network, called a ``layer" of the multi-layer network. Multi-layer networks are a more accurate representation of many complex systems since many entities in those systems are involved simultaneously in multiple networks. This means each of the individual network layers carries only partial information about the interactions among the entities, and full information is available only through the multi-layer networked system \citep{ma11,pc15}. While the term multi-layer networks is often used to include networks with more general connectivity patterns, e.g., inter-layer connections \citep{kivela14}, we restrict our attention to networks with multiple types of relations on the same set of entities and without inter-layer connections. Such networks have also been called ``multiplex networks" in the literature \citep{kivela14}. 

We will consider a number of such inherently multi-layer networks as real world examples in this paper. In the social networking website Twitter, users can engage in various types of interactions with other users, e.g., ``mention", ``follow", ``retweet", etc. \citep{gc13,pc15}. Although the individual network layers created by these relationships are structurally highly related, they represent different facets of user behavior. In another example from biology, the neural network of a small organism, Caenorhabditis elegans, consists of neurons which are connected to each other by two types of connections, a synaptic (electrical) link and an ionic (inter-cellular chemical gap junction) link. The two types of links have markedly different dynamics and hence should be treated as two separate layers of a network instead of fusing together into a single network \citep{boccaletti14}. See \citet{kivela14}, \citet{boccaletti14}, and \citet{nl14} for more examples and discussions of multi-layer networks.

Efficient detection of community structure is an important learning goal in networks. Communities or modules in a network are defined as subsets of nodes which are more similar to each other than the rest of the network. This definition is a bit ambiguous in the sense that a group of nodes might be ``similar" in many different ways. A general acceptable definition is the so called ``structural communities", where nodes within the community are densely connected among themselves as compared to the rest of the network (also known as assortative mixing or homophily \citep{newman03}). The stochastic block model (SBM), a popular model for networks with community structure, generalizes the notion of community to include all nodes, whose connection probabilities to other nodes are similar, in a community (``stochastically equivalent") \citep{hll83,bc09,rcy11}. 
The problem of detecting modules or communities is relatively well studied for single layer networks \citep{f10,gzfa10}. Community detection by optimizing a quality function known as modularity \citep{ng04} is popular in various application areas and its theoretical properties have been studied extensively as well \citep{bc09,zlz12}. However the community detection problem for multi-layer networks has not been studied well in the literature until recently.

Motivated by the goal of developing methods for community detection in multi-layer networks that can take into account the information present in all the network layers simultaneously, we propose a number of null models and derive modularities based on those null models. Recently there has been a surge in analysis of networks with multiple layers \citep{de13,bazzi14, boccaletti14, kivela14,hxa14,pc15,peixoto15,stanley15,taylor15}. A few modularity measures have also been proposed in the literature. \citet{mucha10} used a null model formulated in terms of stability of communities under Laplacian dynamics in networks to derive a modularity measure for multi-layer networks with inter-layer node coupling. Another extension of the Newman-Girvan modularity \citep{ng04} to multi-layer settings as a sum of layer-wise modularities was mentioned by \citet{liu14} and \citet{slcp14}. Aggregation of adjacency matrix as a way of combining information from multiple graphs has also been explored in the literature. However, recent information theoretical results show that community detection on such aggregated graph does not always perform quite well as the different types of layers might have quite different properties in terms of density and signal quality which will get lost due to aggregation \citep{pc15}. In this paper, we are interested in the problem of detecting a common set of communities for the nodes, which is distinct from the problem of detecting different (but perhaps related) communities in the layers of the network. \\

The main contributions of this paper can be summarized as follows.
\begin{itemize}
    \item We propose two sets of null models for multi-layer networks, independent degree (ID) and shared degree (SD) models, depending upon how the degrees of different layers are modeled. We propose simple closed form estimators for the parameters of the models and study their asymptotic consistency properties. We also propose a hypothesis test to select between the two models. 
    \item We use two variations of the above degree based null models, based on whether the null model conditions on expected (unknown) degrees or observed degree sequence, and develop a series of modularity and likelihood quality function measures on the basis of those null models. The new null models can be thought of as multi-layer extensions of the expected degree and configuration models widely employed in community detection. We study the properties and performance of the proposed quality functions and maximization procedures in simulated and real multi-layer networks. 
\end{itemize}

The rest of paper is organized as follows. Section 2 deals with degree distributions in multi-layer networks, defines two families of multi-layer null models,  and develops tools for parameter estimation and model selection. Section 3 extends the Newman-Girvan modularity to multi-layer settings with various multi-layer configuration models as null models. Section 4 first defines four multi-layer degree corrected stochastic block models and then derives modularity measures based on them. Section 5 deals with computational issues, while Section 6 presents a study of the proposed modularities to assess their effectiveness on simulated networks. Section 7 applies the methods for community detection in a number of real world networks and Section 8 gives concluding remarks.

\section{Multi-layer null models: estimation and selection}

We represent an undirected multi-layer graph as $G=\{V,E\}$, where the
vertex set $V$ consists of $N$ vertices representing the entities and the edge set $E$ consists
of edges of $M$ different types representing the different relations.  We define the adjacency matrix $A^{(m)}$ corresponding to the $m$th network layer as follows:
\begin{equation*}
A_{ij}^{(m)}= \begin{cases} 1, & \text{if there is an } m \text{th type of edge between } i \text{ and } j, \\ 0, & \text{otherwise.} \end{cases}
\end{equation*}
Note that we do not consider the possibility of inter-layer edges in our multi-layer network. The  multi-layer network can be viewed as a graph with vector valued edge information with the ``edge-vector" between two nodes $i$ and $j$ being $\mathbf{A}_{ij}=\{A_{ij}^{(1)},A_{ij}^{(2)},$ $\ldots,A_{ij}^{(M)}\}$. In this connection we also define the ``multi-degree" of node $i$ as $\mathbf{k}_{i}=\{k_{i}^{(m)} ;\ m=1, \ldots , M \} $ where $k_{i}^{(m)}=\sum_{j} A_{ij}^{(m)}$ is its degree of $m$th type. For a multi-layer network with $K$ communities, we further denote the community assignment vector of the nodes as $\mathbf{z}=\{z_{i} ;\ i=1, \ldots , N \}$ with $z_{i}$ taking values in the set $\{1, \ldots , K\}$. We will use the notations $e^{(m)}_{ql}$ and $e^{(m)}_q$ to denote the total number of edges of type $m$ between communities $q$ and $l$ and the total degree of type $m$ of all nodes in community $q$, i.e., $e_{ql}^{(m)}= \sum_{i,j} A_{ij}^{(m)} I(z_i=q,z_j=l) $ and $e_{q}^{(m)}=\sum_{i,j} A_{ij}^{(m)} I(z_i=q)= \sum_{i} k_{i}^{(m)} I(z_i=q) $, where $I(\cdot)$ is the indicator function which is 1 if the condition inside is satisfied and 0 otherwise. Note that $e^{(m)}_{ql}=e^{(m)}_{lq}$,  $e_{q}^{(m)} = \sum_{l} e^{(m)}_{ql} $, and $e_{qq}^{(m)}$ is twice the number of edges within the community $q$, for all $m$ in an undirected multi-layer graph. The total number of edges in layer $m$ and in the entire network are denoted as $L^{(m)}$ and $L$ respectively.

\subsection{Null models for community detection in single layer networks}
Modularity can be viewed as a score that computes the difference between the observed structure of the network from an expected structure under a random ``null" network. The null network can be generated by a random network null model which creates connections between nodes at random without any special structure of interest \citep{slcp14,bazzi14}. In particular, for community detection the modularity score computes the difference between the observed number of edges and that expected in a null network within a group of nodes marked as a community. The community structure is strong if this score, with a proper normalization, is high. As an example, the celebrated Newman-Girvan modularity \citep{ng04} for single layer networks has the following expression:
\begin{equation}
Q_{NG} = \frac{1}{2L} \sum_{i,j} \Bigg \{ A_{ij} - \frac{k_{i} k_{j}}{2L}\Bigg \} \delta (z_{i}, z_{j})=\frac{1}{2L} \sum_{q} \Bigg \{e_{qq} - \frac{e^{2}_{q}}{2 L}\Bigg \},  \label{NG}
\end{equation}
where $k_{i}$ and $e_{q}$ are the degree of node $i$ and the total degree of community $q$ respectively, and $e_{qq}$ and $L$ represent twice the total number of edges within community $q$ and the total number of edges in the entire network respectively.

Although originally the modularity score in Equation \ref{NG} was  heuristically motivated, one can recognize the configuration model or a random graph on give degree sequence \citep{bollobas2001random,molloy1995critical} as the null model that generates the null network for this modularity measure. In the configuration model $\mathcal{G} (N,\mathbf{k} )$ for a single layer network with number of nodes $N$ and given degree sequence $\mathbf{k}=\{ k_{i} \}$, the null network is sampled from a population of networks having the same degree sequence through random matching of nodes. For some community assignment of the nodes of the network, the method then computes the expected number of edges according to this null network within each community. The modularity score is then the difference between the observed number of edges and the expected number of edges obtained in the previous step. Optimizing this modularity score across all possible community assignments will then lead to a detection strategy for network communities.

Another related model is the ``expected degree model" by \citet{chunglu02}, which can be thought of as a null model for a likelihood modularity based on the degree corrected block model \citep{zlz12}. In this model each vertex $i$ is associated with a parameter $w_i$ which represents its expected degree. The probability $P_{ij}$ of an edge between nodes $i$ and $j$ is proportional to the product of the expected degrees $\kappa_i$ and $\kappa_j$ \citep{chung2002average}:
\begin{equation}
P_{ij} = \frac{\kappa_i \kappa_j}{\sum_{k} \kappa_k}, \quad \max_{i} \kappa_i^2 \leq \sum_{k} \kappa_k.
\label{CLmodel}
\end{equation}
The null network is then formed by adding edges $A_{ij}$ between nodes $i$ and $j$ independently with probability $P_{ij}$, i.e., $A_{ij} \sim Bernoulli(P_{ij})$. For the subsequent discussions we use the re-parameterization in \citet{pw12} and \citet{arcolano2012moments}:
 \begin{equation}
A_{ij} \sim Bernoulli (P_{ij}), \quad \quad P_{ij} = \theta_i \theta_j, \quad \quad \theta_i \in (0,1)
\label{expected_degree_model}.
\end{equation}
Note the model includes the possibility of self-loops with probability of a self-loop for node $i$ being $\theta_i^2$.

Let $\kappa_i$ denote the expected degree of a node and $2\mathcal{L} = \sum_i \kappa_i$ denote twice the expected number of edges. We note that under this model, the expected degree of a node and twice the expected number of edges are
\[
\kappa_i =  \theta_i \sum_j \theta_j, \quad \quad 2\mathcal{L} = \sum_i \sum_j \theta_i \theta_j = (\sum_k \theta_k)^2,
\]
so that 
\[
\frac{\kappa_i}{\sqrt{2\mathcal{L}}} = \theta_i.
\]

This expression for $\theta$ motivates a commonly used estimator for estimating $\theta$ \citep{pw12,olhede2012degree,arcolano2012moments}
\begin{equation}
 \hat{\theta}_{i}=\frac{k_{i}}{\sqrt{2L}}. \label{approxMLE}
\end{equation}
Statistical properties of the estimator have been studied previously in the literature under varying assumptions. \citet{arcolano2012moments} derived moments of the estimator without any additional assumptions. Asymptotic normality of the estimator was studied under the assumption of decaying tail degree similar to power law degree distribution in \cite{olhede2012degree}. In this paper we assume a condition on the average expected degree and prove that the estimator is consistent for estimating $\theta_i$ in Thoerem 1. Our prove technique is different from that of \cite{olhede2012degree}.

A consequence of the estimator is that the estimated probability of a link between $i$ and $j$ is given by $\hat{P}_{ij}=\frac{k_{i} k_{j} }{2L}$, which is the same as the one obtained from the configuration model \citep{pw12}.  The authors in \citet{pw12} further gave a motivation for the estimator as MLE when the Bernoulli distribution is approximated with the  Poisson distribution for ease of computation (the model continues to allow self-loops). The existence and consistency of MLE in the Bernoulli model with logit links has been analyzed in \cite{chatterjee2011random} and \cite{rinaldo2013maximum}.

The Poisson distributed version of the random graph has been used previously in the literature as null model for community detection \citep{kn11,zlz12, pw12, yan14}. Usually the distribution leads to multiple edges and the random variable $A_{ij}$ represents the count of the edges between $i$ and $j$, while $P_{ij}$ represents the expected number of such edges. It has been argued that often graphs naturally contain multiple edges (weighted graphs) and then the Poisson model is appropriate. However, the Poisson distribution is also valid as an approximation to the binary graph since we expect $P_{ij}$ to be small in modern large scale networks which are quite sparse, and in such cases both distributions would lead to similar results \citep{zlz12, pw12, yan14}. This can be easily seen by comparing the moment generating functions (MGFs) of the two distributions:
\begin{align*}
M_{Poi}(t)  =\exp (p(e^t-1)), \quad \quad 
M_{Bern}(t)  =1 -p + pe^t.
\end{align*}
A first order Taylor expansion of the MGF of Poisson distribution shows
\[
M_{Poi}(t) =1 + p (e^t -1) + O (p^2 (e^t -1)^2).
\]
Therefore, if $p \rightarrow 0$, then the MGF of Poisson converges to the MGF of Bernoulli.

Even though the estimator in Equation (\ref{approxMLE}) is derived assuming that the network contains self-loops, the estimator can be used for networks without self-loops as well. \citet{pw12} showed that this estimator, when plugged into the likelihood equation of a graph with Poisson distributed edges but without self-loops, gives only a small error. They further showed that the solution is a good approximation for the MLE in the original Bernoulli model (without self-loops) as well.

Nevertheless, we continue to consider the settings of the model in Equation (\ref{expected_degree_model}), i.e., we have an unweighted random graph with edges distributed according to Bernoulli distribution and contains self-loops. We prove uniform consistency of the estimator in Equation (\ref{approxMLE}) in the following theorem. Define the average expected degree of $G$, $\bar{\kappa} =\frac{2\mathcal{L}}{N} = \frac{1}{N}(\sum_k \theta_k)^2$.

\begin{thm}
Let $G$ be a graph generated according to the expected degree model in Equation (\ref{expected_degree_model}) with parameters $\bar{\theta}_1,\ldots, \bar{\theta}_N$. Assume the average expected degree of $G$, $\bar{\kappa} \geq C \log N$, for a sufficiently large constant $C$ not dependent on $N$. Then the parameter estimates in Equation (\ref{approxMLE}) satisfy
\[
\sup_{i \in \{1,\ldots,N\}}|\hat{\theta}_i -\bar{\theta}_i| \overset{p}{\to} 0, \text{ as } N \to \infty.
\]
\end{thm}

The proof of this theorem and two other theorems later are in the Appendix. We have assumed that the average expected degree grows at a rate faster than $\log N$. This is a stronger condition than that in \cite{olhede2012degree}, however unlike \cite{olhede2012degree}, we do not require the degree distribution to decay.

\subsection{Degrees in multi-layer network and null models}

While the observed number of edges among the nodes within a community is unique, its expectation can vary depending on which network null model is chosen. Hence there can be considerable variation in the communities detected using a modularity score. The null model should be wisely chosen with the aim to capture all sources of systematic variation in the network except the community structure. So given that the observed network is a realization of the null model, the additional edges observed within the communities beyond what is expected from a purely random phenomenon should be attributed to the community structure. 

From the preceding discussions it is clear that the degree sequence, observed or expected, plays a major role in null models. In a multi-layer network, every node is associated with a ``multi-degree" vector instead of a single degree. Hence in multi-layer networks degree heterogeneity might be present in two aspects: across the nodes in a layer and across layers in a node. The across layer heterogeneity can be due to two reasons. First, some layers might be inherently sparse and some might be dense, and second, nodes might participate in varying degrees in relations captured by different layers. To illustrate this, consider the British Members of Parliaments (MPs) in twitter dataset  \citep{gc13,pc15}. While there is clear degree heterogeneity across the different MPs within a network layer depending on their political influence and significance, there might also be degree heterogeneity across layers for the same MP depending upon personal preferences. Moreover, the layer ``follow" is somewhat denser compared to the layers ``mention" and ``retweet", possibly because the former requires one time attention and the later, continued. Hence a number of null models are possible depending upon how one models the degree sequence.

\begin{figure}[h]
\centering{}
\begin{subfigure}{0.3\textwidth}
\centering{}
\includegraphics[width=.95\linewidth]{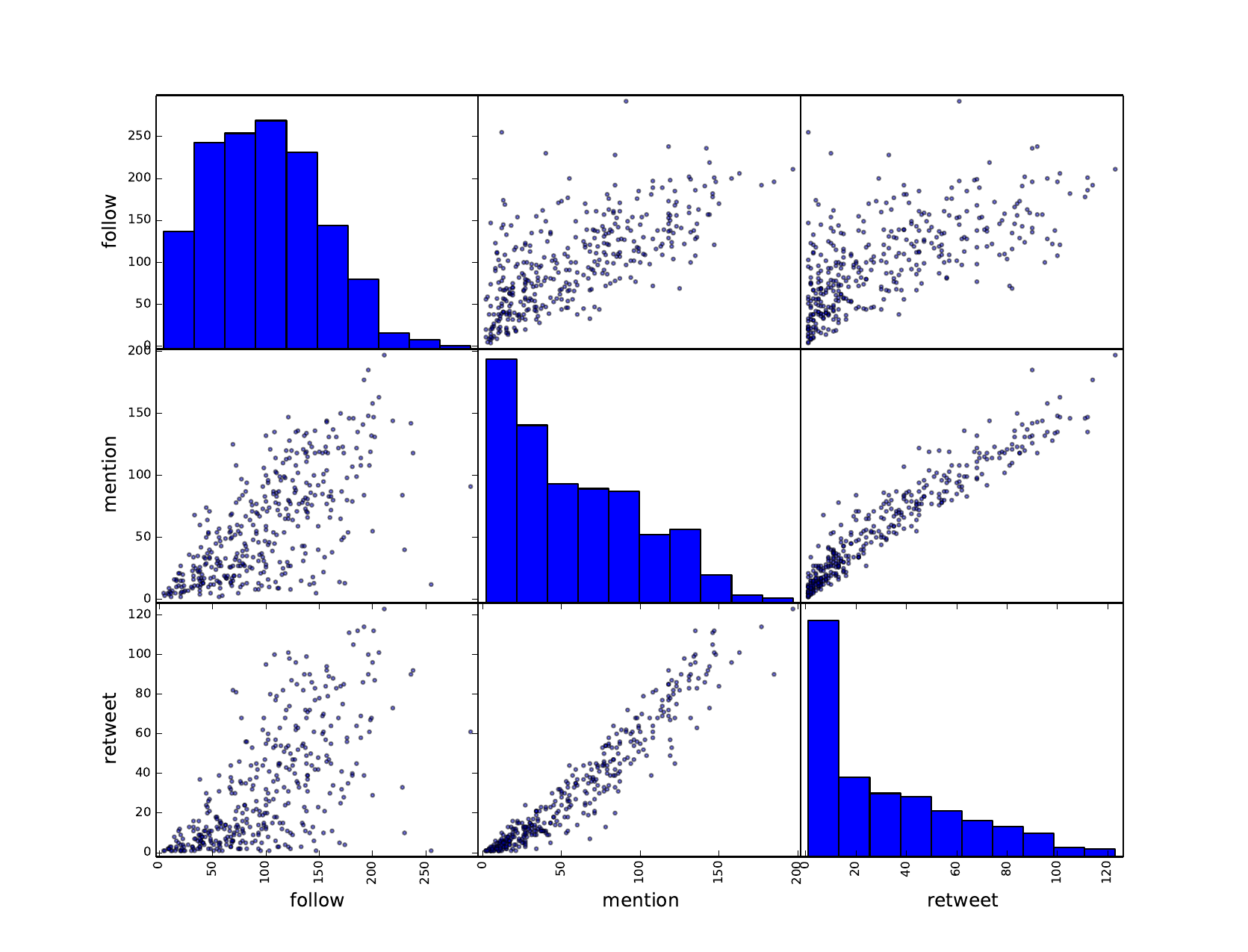}
\end{subfigure}%
\begin{subfigure}{0.3\textwidth}
\centering{}
\includegraphics[width=.95\linewidth]{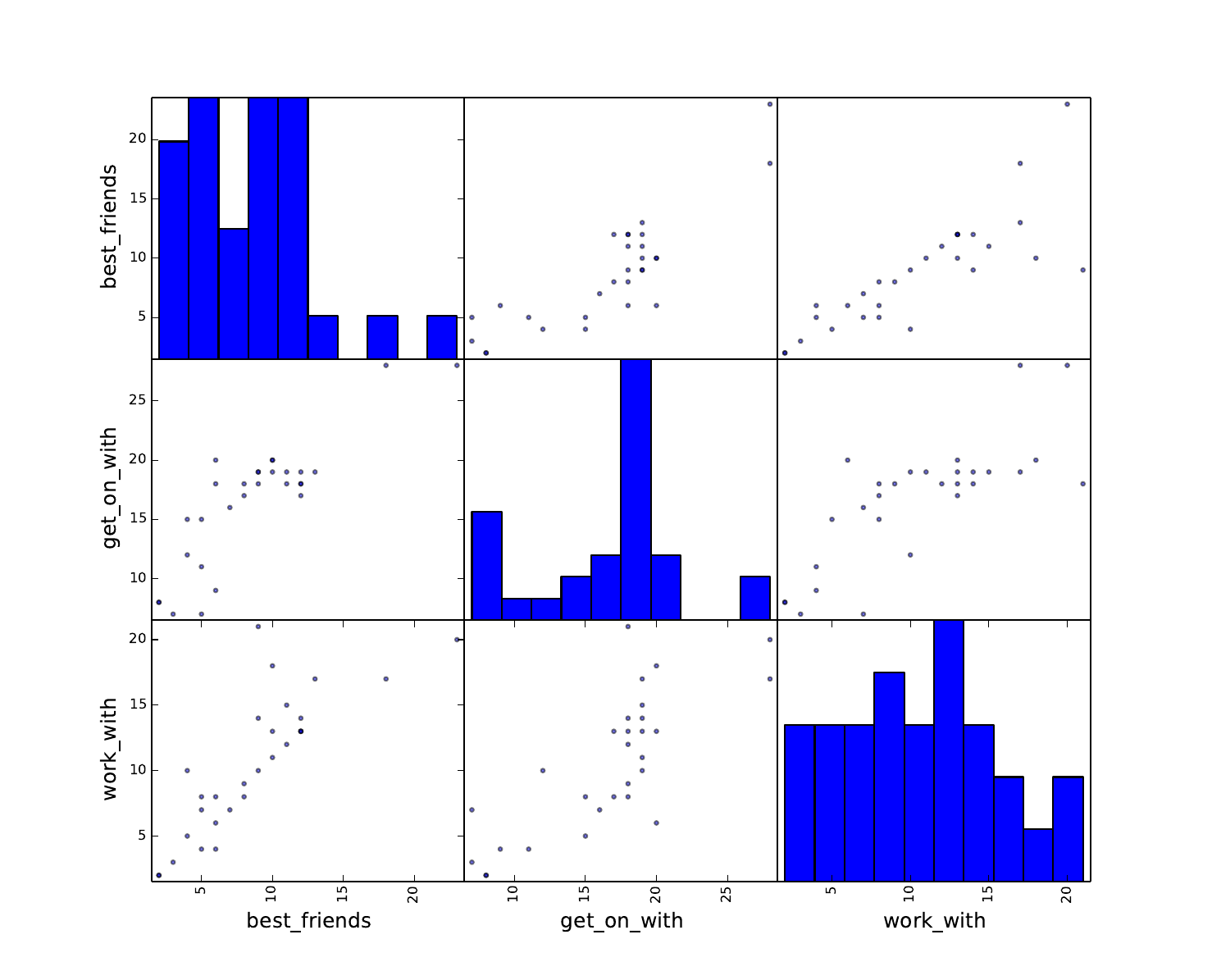}
\end{subfigure}%
\begin{subfigure}{0.3\textwidth}
\centering{}
\includegraphics[width=.95\linewidth]{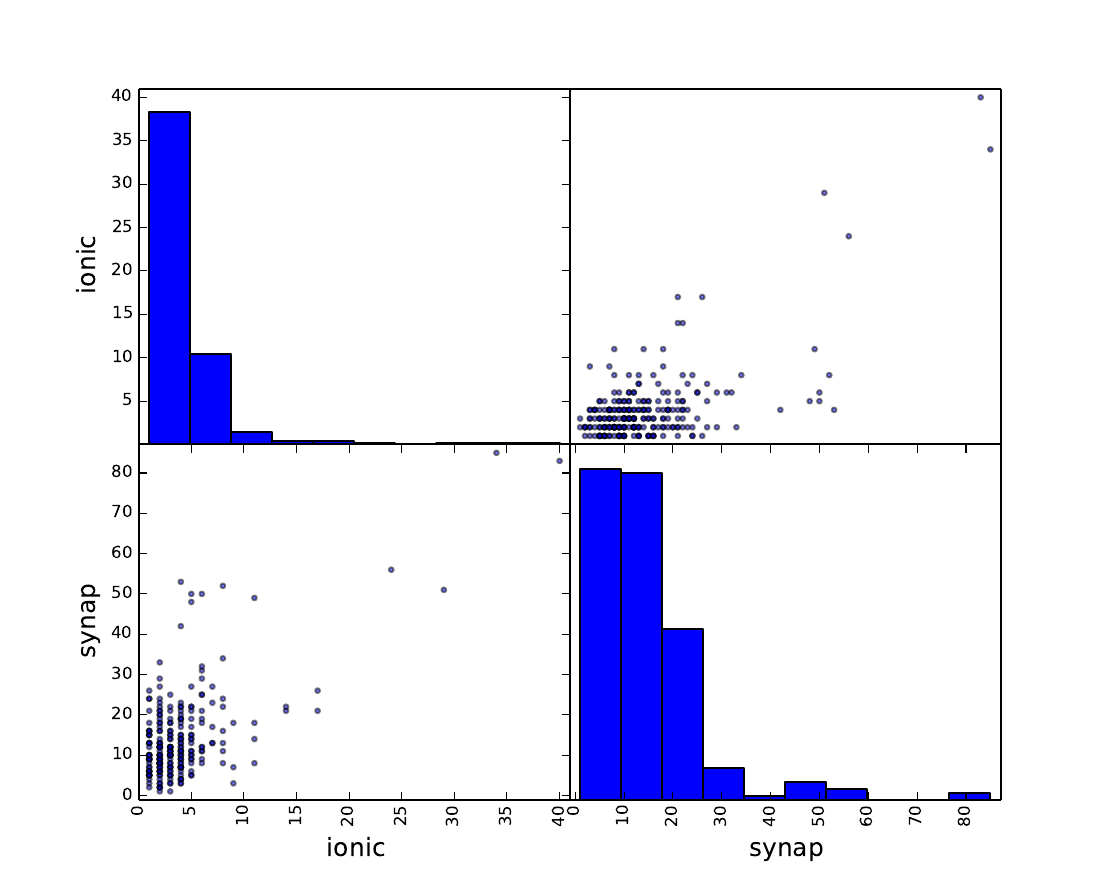}
\end{subfigure}
\begin{center} (a1) \hspace{100pt} (b1) \hspace{100pt} (c1) \end{center}
\begin{subfigure}{0.22\textwidth}
\centering{}
\includegraphics[width=.95\linewidth]{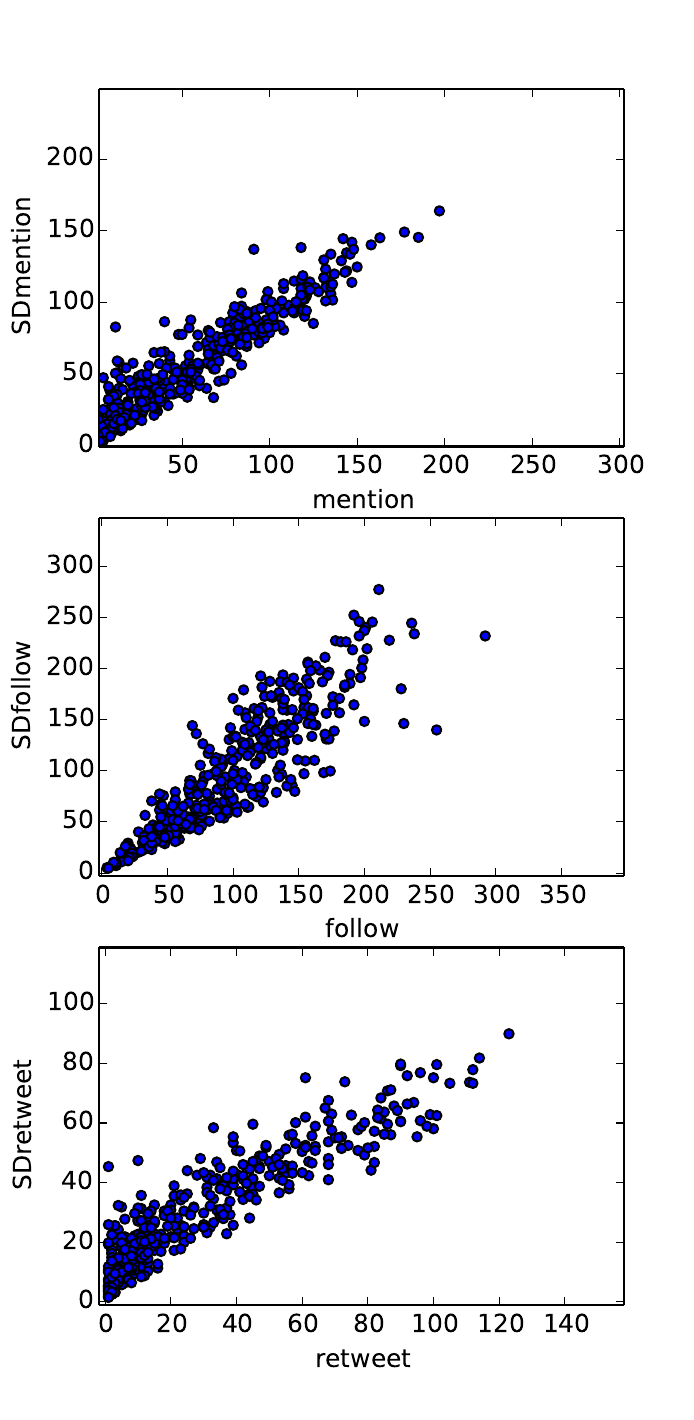}
\end{subfigure}%
\hspace{20pt}
\begin{subfigure}{0.22\textwidth}
\centering{}
\includegraphics[width=.95\linewidth]{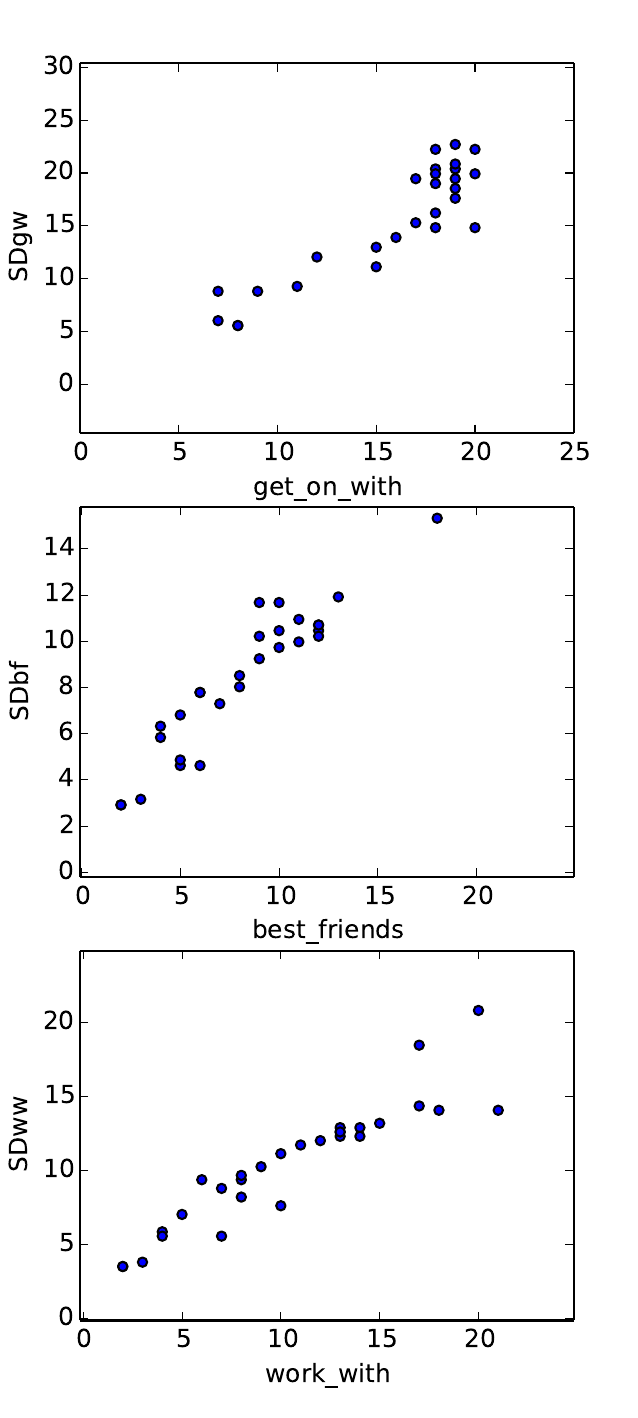}
\end{subfigure}%
\hspace{20pt}
\begin{subfigure}{0.22\textwidth}
\centering{}
\includegraphics[width=.95\linewidth]{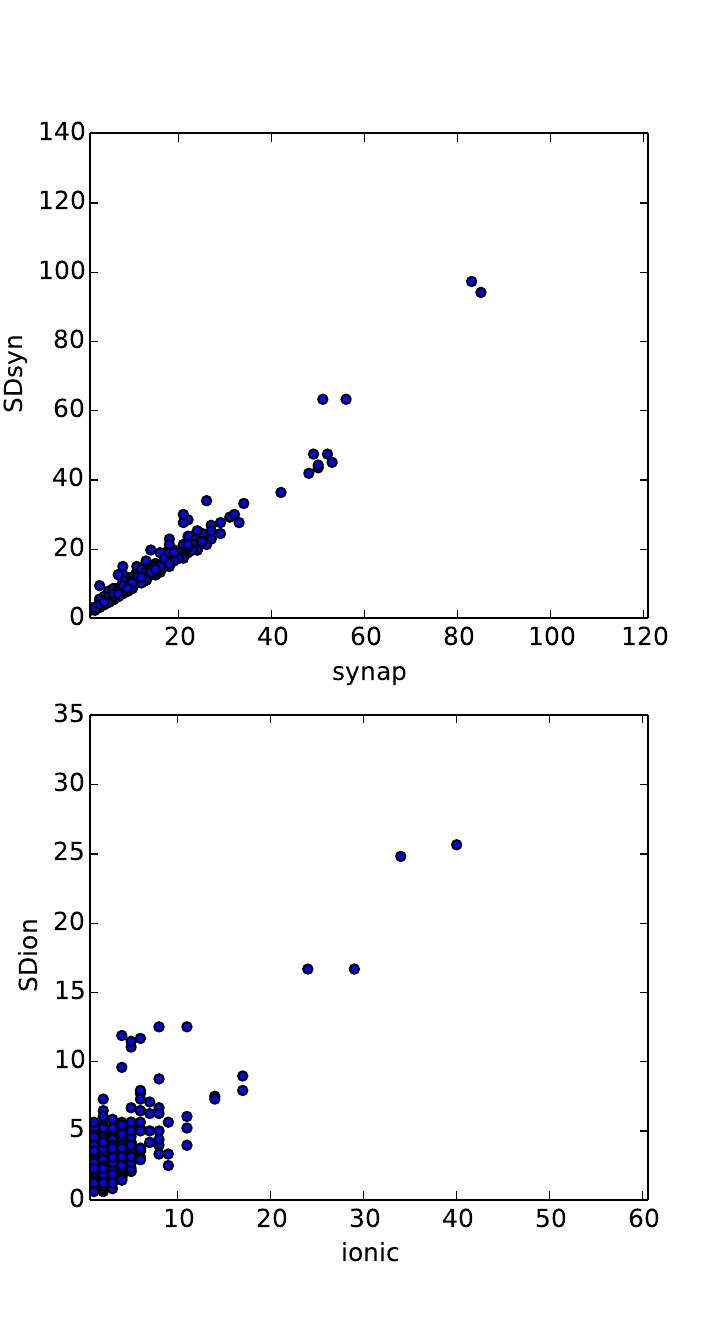}
\end{subfigure}
\begin{center} (a2) \hspace{100pt} (b2) \hspace{100pt} (c2) \end{center}
\caption{Observed degree distributions across layers in (a1) Twitter network of British MPs, (b1) 7th grade school students and (c1) C-elegans neuronal network; (a2, b2, c2) degree distribution fitted with a shared degree model plotted as scatter plots with the observed degrees in each layer.} \label{fig:degree}
\end{figure}

We can broadly classify our null models into independent degree (ID) models and shared degree (SD) models. The independent degree models assume degrees in each layer are independent of the degrees in other layers, and assign a separate degree parameter for each layer to each node.  Hence we can write the independent degree multi-layer (expected degree) model as
 \begin{equation}
A^{(m)}_{ij} \sim Bernoulli (P^{(m)}_{ij}), \quad \quad P^{(m)}_{ij} = \theta^{(m)}_i \theta^{(m)}_j, \label{IDexpected}
\end{equation}
In contrast, the shared degree models assign only one degree parameter to each node, and layer-wise variations in degree are captured by a single layer-specific global parameter. The shared degree multi-layer (expected degree) model can be written as
 \begin{equation}
A^{(m)}_{ij} \sim Bernoulli (P^{(m)}_{ij}), \quad \quad P^{(m)}_{ij} = \theta_i \theta_j \beta_{m}. \label{SDexpected}
\end{equation}
The shared degree model further requires an identifiability constraint $\sum_{m} \beta_{m}=1$. Note that the ID model requires $MN$ parameters while the SD model requires only $N+M-1$ parameters. Since we have $M$ network layers, we effectively have $O(MN^2)$ data points. Hence in the context of sparse individual layers, the SD model, being more parsimonious, might lead to less variance at the expense of bias.

It has been empirically observed that the layers of a multi-layer network have many structural similarity \citep{kivela14}. Among others, it has been shown  that the degrees are often highly correlated \citep{nl14}. Since we expect the individual layers to be manifestation of an underlying common structure, the degrees of a node in different layers are also expected to be highly inter-related. While there are instances where the degrees can be negatively correlated, a large number of  cases have positively correlated degrees. Figures \ref{fig:degree}(a1), (b1) and (c1) show the degree distribution of three layers (mention, follow, retweet) in Twitter network of British MPs, three layers (get on with, best friends, work with) in a friendship network of 7th grade school students, and two layers (synaptic, ionic) of the neuronal network of C-elegans respectively. Three real-world multi-layer networks from diverse fields, ranging from social networks, friendship networks to neurological networks, exhibit a large positive correlation in the degree distribution among their layers. In such cases a relatively parsimonious shared degree null model described in (\ref{SDexpected}) might be appropriate. Figures \ref{fig:degree}(a2), (b2), and (c2) show that the shared degree null model fits well (using the parameter estimators that we describe in the next section) to the degree distribution of the layers in these networks respectively.

\subsection{Estimation in multilayer ID and SD models}
To derive parameter estimates for the multilayer ID and SD models in Equations (\ref{IDexpected}) and (\ref{SDexpected}) respectively, once again we approximate the Bernoulli log-likelihood with the Poisson log-likelihood. We justify the resulting estimators by showing that they are consistent estimators for the parameters.

The Poisson approximated log-likelihood from the ID model (\ref{IDexpected}) can be written as
\begin{equation*}
l(\theta) =\sum_{m}\sum_{i,j}\{A_{ij}^{(m)} (\log(\theta^{(m)}_{i})+\log(\theta^{(m)}_{j})) -\theta^{(m)}_{i}\theta^{(m)}_{j} \} - \sum_{m}\sum_{i,j} \log (A^{(m)}_{ij} !).
\end{equation*}
The likelihood equations are therefore,
\[
\frac{\partial l }{\partial \theta^{(m)}_i}: \ \  \frac{\sum_j A_{ij}^{(m)}}{\theta^{(m)}_i} - \sum_{j } \theta_j^{(m)} = 0, \quad \quad i = \{1,\ldots N\}, \quad m=\{1,\ldots, M\}.
\]
A solution to the likelihood equation leads to the following simple estimator:
\begin{equation}
\hat{\theta}^{(m)}_{i} = \frac{k_{i}^{(m)}}{\sqrt{2L^{(m)}}},    \label{MLE_ID}
\end{equation}
which can be thought of as the multi-layer extension of the estimator in  Equation (\ref{approxMLE}).


For the SD model the Poisson approximated likelihood is
\begin{equation*}
l(\theta,\beta) =\sum_{m}\sum_{i,j}\{A_{ij}^{(m)} (\log(\theta_{i})+\log(\theta_{j}) + \log (\beta_m)) -\theta_{i}\theta_{j}\beta_m \} - \sum_{m}\sum_{i,j} \log (A^{(m)}_{ij} !).
\end{equation*}
Therefore the set of likelihood equations is given by
\begin{align*}
    \frac{\partial l }{\partial \theta_i} & :\ \  \frac{\sum_m \sum_j A_{ij}^{(m)}}{\theta_i} - \sum_m\sum_{j } \theta_j = 0, \quad \quad i = \{1,\ldots N\}, \\
    \frac{\partial l }{\partial \beta_m} & :\ \  \frac{\sum_{i,j} A_{ij}^{(m)}}{\beta_m} - \sum_{i,j } \theta_i\theta_j = 0, \quad \quad  m=\{1,\ldots, M\}.
\end{align*}
Solving the likelihood equations, we obtain the following estimators:
\begin{equation}
    \hat{\theta}_{i}=\frac{\sum_{m} k_{i}^{(m)}}{\sqrt{2L}}, \quad \quad  \hat{\beta}_{m}=\frac{L^{(m)}}{L}.
    \label{MLE_SD}
\end{equation}

The parameter estimates are uniformly consistent under the original Bernoulli model which is given by the following two theorems. We start with a few observations that serve as motivation for the estimators in Equations (\ref{MLE_ID}) and (\ref{MLE_SD}). Let $\kappa_i^{(m)}$ denote the expected degree of a node $i$ in the $m$th layer and $2\mathcal{L}^{(m)} = \sum_i \kappa_i^{(m)}$ denote twice the expected number of edges in layer $m$. Further, let $\mathcal{L} = \sum_m \mathcal{L}^{(m)}$ be the total number of edges in the multi-layer network. We note that under the ID model,
\[
\kappa_i^{(m)} =  \theta_i^{(m)} \sum_j \theta_j^{(m)}, \quad \quad 2\mathcal{L}^{(m)} = \sum_i \sum_j \theta_i^{(m)} \theta_j^{(m)} = (\sum_k \theta_k^{(m)})^2,
\]
and consequently, $
\frac{\kappa_i^{(m)}}{\sqrt{2\mathcal{L}^{(m)}}} = \theta_i^{(m)}.$ 
Similarly, for the SD model
\[
\sum_m\kappa_i^{(m)} =  \theta_i \sum_j \theta_j \sum_m \beta_m = \theta_i \sum_j \theta_j, \quad \quad 2\mathcal{L}^{(m)} = \sum_i \sum_j \theta_i \theta_j \beta_m = \beta_m (\sum_k \theta_k)^2,
\]
\[
2\mathcal{L} = \sum_i \sum_j \sum_m \theta_i \theta_j \beta_m = (\sum_m \beta_m) (\sum_k \theta_k)^2 =(\sum_k \theta_k)^2  .
\]
Note that here we have used the model constraint that $\sum_m \beta_m =1$. Therefore, for the SD model, 
\[
\frac{\sum_m \kappa_i^{(m)}}{\sqrt{2\mathcal{L}}} = \theta_i, \quad \quad \frac{\mathcal{L}^{(m)}}{\mathcal{L}} = \beta_m.
\]
\begin{thm}
Let $G$ be a multi-layer graph generated according to the independent degree model in Equation (\ref{IDexpected}) with parameters $\bar{\theta}_i^{(m)}, \, i=\{1,\ldots,N\},\  m=\{1,\ldots, M\}$. Assume  $\mathcal{L'} =\min_m \mathcal{L}^{(m)} \geq C N \log (MN)$, for a sufficiently large constant $C$ not dependent on $N$. Then the parameter estimates in Equation (\ref{MLE_ID}) satisfy
\[
\sup_{\substack{i \in \{1,\ldots,N\}, \\
    m \in \{1,\ldots,M\}}}|\hat{\theta}_i^{(m)} -\bar{\theta}_i^{(m)}|  \overset{p}{\to} 0, \text{ as } N \to \infty.
\]
\end{thm}

\begin{thm}
Let $G$ be a multi-layer graph generated according to the shared degree model in Equation (\ref{SDexpected}) with $\bar{\theta}_1,\ldots, \bar{\theta}_N, \bar{\beta}_1,\ldots, \bar{\beta}_M$. Assume  $\bar{\mathcal{L}} = \frac{1}{M}\sum_m \mathcal{L}^{(m)} \geq C N \log N$, for a sufficiently large constant $C$ not dependent on $N$. Then the parameter estimates in Equation (\ref{MLE_SD}) satisfy 
\begin{align*}
\sup_{i \in \{1,\ldots,N\}}|\hat{\theta}_i -\bar{\theta}_i|  & \overset{p}{\to} 0, \text{ as } N \to \infty, \\
\sup_{m \in \{1,\ldots,M\}}|\hat{\beta}_m -\bar{\beta}_m|  & \overset{p}{\to} 0, \text{ as } N \to \infty.
\end{align*}
\end{thm}

The proofs of these two theorems are in the Appendix. We note that the condition required on the expected number of edges of the multi-layer network is stronger for Theorem 2 than for Theorem 3. While in Theorem 2 on ID model we need the minimum expected degree across layers to be $O(N \log (MN)$, for Theorem 3 on SD model we need the average expected degree across layers to be $O(N \log N)$.

Finally, throughout this section we have allowed the possibility of self-loops in the models. However, many observed networks do not contain self-loops. Therefore, we also study the effect of not allowing for self-loops on the estimators and how much error the estimators make in that situation in the Appendix.

We summarize the properties of the estimators below.
\begin{itemize}
    \item The estimators are consistent under the original Bernoulli model even though they are not the MLE under that model. We prove this result directly through concentration inequalities. 
    \item The estimators are MLE under the Poisson approximation model.
    \item If the model does not allow self-loops, the estimators are not MLE even under the Poisson approximation. We do not have any consistency results under this setting, however, we discuss the extent of expected difference of the estimators from the MLE under the model.
\end{itemize}

\subsection{Null model selection}
\label{sec:modelselection}
By dissociating the degree based null model from the community structure component of a modularity measure, we make it easier to first choose the appropriate null model based on observed degree pattern, and then choose an appropriate modularity measure based on the null model. In this context a question that naturally arises is, given a multi-layer network how would one choose between an independent degree and a shared degree null model?

A hypothesis testing based framework for model selection was developed in \citet{yan14} for selecting between the ordinary stochastic blockmodel (SBM) and the degree corrected block model (DCBM) in single layer networks. In our case, however, the question is not between choosing degree correction or not, rather between what kind of degree correction is required; an independent degree model or a shared degree model. Here we provide a guidance through a simple approximate model selection procedure based on likelihood ratio calculations. The null hypothesis is that the SD model is the true data generating model whereas the alternative hypothesis is that it is the ID model that generates the data. 
The maximized Poisson log-likelihood with the MLE solution as in (\ref{MLE_ID}) is
\begin{equation}
\Lambda_1= \sum_{i} \sum_{m} k_{i}^{(m)} \log \frac{k_{i}^{(m)}}{\sqrt{2L^{(m)}}}  - L + c_{1}, \label{Lambda1}
\end{equation}
where $c_{1}=\sum_{m}\sum_{i<j} \log (A^{(m)}_{ij} !)$. For the SD model, using the MLE in Equation (\ref{MLE_SD}), the maximized Poisson log-likelihood is
\begin{equation}
\Lambda_2= \sum_{i} \sum_{m} k_{i}^{(m)} \log \frac{\sum_{m} k_{i}^{(m)}}{\sqrt{2L}}+ \sum_{m} L^{(m)} \log \left(\frac{L^{(m)}}{L}\right)  - L + c_{1}. \label{Lambda2}
\end{equation}
Note that $c_{1}$ and the $L$ terms cancel when we subtract (\ref{Lambda2}) from (\ref{Lambda1}) to compute the logarithm of likelihood ratio. 
The standard theory on likelihood ratio tests would suggest that $2(\Lambda_1 -\Lambda_2)$ is distributed as a $\chi^2$ distribution with degrees of freedom $MN-(N+M-1)$. However there is some concern about the validity of the assumptions under which the asymptotic distribution of the test statistic is usually derived in the present case. In particular the ``effective sample size" in sparse multi-layer graphs (average degree per layer is $O(1)$) is $O(MN)$ and the ID model contains $MN$ parameters, leading to the failure of standard asymptotics \citep{yan14}. In Theorems 2 and 3 we have shown consisteny of the MLEs. However, it is not clear immediately if the asymptotic normality with $\sqrt{n}$ rate holds for the MLEs and consequently if the second order delta method can be applied.  Hence we use parametric bootstrap to compute the empirical distribution of the likelihood ratio test statistic under the null model. In particular we fit the SD model to the data and estimate the parameters $\hat{\theta}_{i}, \hat{\beta}_{m}$. We then generate a large number of networks (we used 1000 in data analysis) from the SD model with the estimated parameters and compute our test statistic on each of those networks. The values of the statistic form an empirical distribution which is subsequently used to calculate $p$-value for the test.
Once a null model selection is performed, the user can choose an appropriate modularity measure among the ones we define in upcoming sections.

\section{Multi-layer configuration models and modularity measures}

Similar to the multi-layer extensions of the expected degree model developed in the previous section, we define several extensions of the configuration model for multi-layer networks, conditioned upon the observed multi-degree sequence $\mathbf{k}=\{\mathbf{k}_1,\ldots,\mathbf{k}_N\}$.

In the first model we propose, we assume the degree sequence in one layer is independent of the degree sequence in other layers.  The number of stubs or half-edges (one end of an edge) coming out of node $i$ is $k_{i}^{(m)}$. For every stub, there are $2L^{(m)}-1$ stubs or half-edges available to connect to. Out of these half-edges, the number of half-edges that will lead to an edge of type $m$  between nodes $i$ and $j$ is $k_{j}^{(m)}$. Therefore, the probability of a connection of type $m$ between nodes $i$ and $j$ is given by $\frac{k_{i}^{(m)}k_{j}^{(m)}}{2L^{(m)}-1}\approx \frac{k_{i}^{(m)}k_{j}^{(m)}}{2L^{(m)}}$, which is the same as the estimate of $p_{ij}$ in the ID model in Equation (\ref{IDexpected}) using the estimator in Equation (\ref{MLE_ID}).  It is a general convention in configuration model to write $2L^{(m)}$ in the denominator instead of $2L^{(m)}-1$ for simplification as $L^{(m)}$ is generally quite large. We call this model the independent degree multi-layer configuration model (ID-MLCM). Using this model as a null model, we then define our first extension of Newman-Girvan (NG) modularity, which we call the multi-normalized average (MNavrg) since the expression is effectively an average of the layer-wise normalized NG modularities:
 \begin{align}
Q_{MNavrg} =\frac{1}{M}\sum_{m} \sum_{i,j} \frac{1}{2L^{(m)}}\Bigg \{ A_{ij}^{(m)} - \frac{k_{i}^{(m)} k_{j}^{(m)}}{2 L^{(m)}}\Bigg \} \delta (z_{i},z_{j}) 
=\frac{1}{M}\sum_{m} \sum_{q} \frac{1}{2L^{(m)}}\Bigg \{ e_{qq}^{(m)} - \frac{(e_{q}^{(m)})^2}{2 L^{(m)}}\Bigg \}.
\label{MNavrg}
\end{align}
There is a similar version of this modularity that has appeared in the literature before for community detection in multi-layer networks --- the intra-layer part of the multi-slice modularity  in \citet{mucha10} and \citet{bassett13} (see also the discussion in \citet{slcp14}), and ``composite modularity" in \citet{liu14}. In those earlier instances, the layer-wise modularity scores are typically not normalized before adding together. To minimize the impact of varying sparsity across layers, it is prudent to normalize the measures by the density of  layers before aggregating \citep{pc15}.

We next describe the shared degree multi-layer configuration model (SD-MLCM). We bring a regularization effect into the null model by sharing degree across layers for networks. This will be particularly useful in the case when all the network layers are extremely sparse. If we do not distinguish the stubs in terms of type, then according to the simple configuration model, the probability of an edge (of any type) between nodes $i$ and $j$ will be given by $\frac{(\sum_{m} k_{i}^{(m)}) (\sum_{m} k_{j}^{(m)})}{2L-1}$. Now, given that these two nodes $i$ and $j$ are the endpoints of a randomly chosen edge, we look into the probability that an edge of type $m$ is formed between the nodes. This probability can be modeled in three different ways, leading to three different shared degree multi-layer configuration models (SD-MLCMs) and consequently three different modularities. To illustrate this, we abuse the notation a little bit and write the probability of an $m$th type of connection between nodes $i$ and $j$ as $P(i,j,m)=P(i,j)\times P(m|i,j)$, where $P(i,j)$ denotes the unconditional probability of an edge between nodes $i$ and $j$, and $P(m|i,j)$ denotes the conditional probability of an $m$th type of edge between $i$ and $j$ given there is an edge between them.

We can use the global frequency of the occurrence of the $m$th type of edge among the multi-layer network as an estimate of the probability $P(m|i,j)$. We call the resulting modularity score shared degree average (SDavrg) since we are using a global estimate for each node. The modularity can be written as
\begin{align}
Q_{SDavrg} &=\frac{1}{M}\sum_{m} \sum_{i,j} \frac{1}{2L^{(m)}} \Bigg \{ A_{ij}^{(m)} - \frac{L^{(m)}(\sum_{m} k_{i}^{(m)})( \sum_{m} k_{j}^{(m)})}{2 L^{2}}\Bigg \} \delta (z_{i}, z_{j}) \nonumber \\
&=\frac{1}{M}\sum_{m} \sum_{q} \frac{1}{2L^{(m)}} \Bigg \{ e_{qq}^{(m)} - \frac{L^{(m)}(\sum_{m} e_{q}^{(m)})^2}{2 L^{2}}\Bigg \}.
\label{SDavrg}
\end{align}

We can also use local estimates of this probability of the $m$th type of connection that is specific to a node or a group of nodes. One such measure would be the ratio of the $m$th type of stubs to all stubs in the communities to which nodes $i$ and $j$ belong. Instead of looking into the entire network, this ratio measures the relative frequency of the occurrence of the $m$th type of edge involving stubs emanating from the groups of either $i$ or $j$. Hence this is a more local measure of the relative density of the $m$th type of edges. The total number of edges (of any type) that have an end in the group to which $i$ and $j$ belong is $\sum_{m} (e^{(m)}_{z_i}+e^{(m)}_{z_j})$. Out of these only $(e_{z_i}^{(m)}+e_{z_j}^{(m)})$ are of type $m$. Hence according to this estimate, the probability of an $m$th type of stub emanating out of $i$ or $j$ is $\frac{e_{z_i}^{(m)}+e_{z_j}^{(m)}}{\sum_{m} (e_{z_i}^{(m)}+e_{z_j}^{(m)})}$. The corresponding modularity, which we call shared degree local (SDlocal) to highlight the fact that it uses a more local estimate of the expected number of edges, is given by
\begin{align}
Q_{SDlocal} &=\frac{1}{M}\sum_{m} \sum_{i,j} \frac{1}{2L^{(m)}} \Bigg \{ A_{ij}^{(m)} - \frac{(e_{z_i}^{(m)}+e_{z_j}^{(m)}) (\sum_{m} k_{i}^{(m)}) (\sum_{m} k_{j}^{(m)})}{\sum_{m} (e_{z_i}^{(m)}+e_{z_j}^{(m)}) 2 L}\Bigg \} \delta (z_{i}, z_{j})  \nonumber \\
& = \frac{1}{M}\sum_{m} \sum_{q} \frac{1}{2L^{(m)}} \Bigg \{ e_{qq}^{(m)} - \frac{e_{q}^{(m)} (\sum_{m} e_{q}^{(m)})}{2 L}\Bigg \}.
\label{SDlocal}
\end{align}

The last one is also a local measure of $P(m|i,j)$, but a more direct measure. We take the ratio of the expected number of edges of type $m$ to the total number of expected edges between the groups to which $i$ and $j$ belong according to the configuration model. Clearly, as per the single layer configuration model, the expected number of edges of type $m$ between groups $z_i$ and $z_j$ is $e_{z_i}^{(m)}e_{z_j}^{(m)}/2L$. Hence in the multi-layer context, given that there is an edge between the groups $z_i$ and $z_j$, the probability that the edge would be of type $m$ is given by $\frac{e_{z_i}^{(m)}e_{z_j}^{(m)}}{\sum_{m} (e_{z_i}^{(m)}e_{z_j}^{(m)})}$. We call this modularity the shared degree ratio (SDratio) to highlight the fact that it takes the ratio of the expected number of edges of type $m$ to the total expected number of edges. The expression is as follows,
\begin{align}
Q_{SDratio} &=\frac{1}{M}\sum_{m} \sum_{i,j} \frac{1}{2L^{(m)}} \Bigg \{ A_{ij}^{(m)} - \frac{(e_{z_i}^{(m)} e_{z_j}^{(m)}) (\sum_{m} k_{i}^{(m)}) (\sum_{m} k_{j}^{(m)})}{\sum_{m} (e_{z_i}^{(m)} e_{z_j}^{(m)}) 2 L}\Bigg \} \delta (z_{i}, z_{j})  \nonumber \\
&=\frac{1}{M} \sum_{m} \sum_{q} \frac{1}{2L^{(m)}} \Bigg \{ e_{qq}^{(m)} - \frac{(e_{q}^{(m)})^2 (\sum_{m} e_{q}^{(m)})^2}{2L \sum_{m} (e_{q}^{(m)})^2 }\Bigg \}.
\label{SDratio}
\end{align}

\section{Degree corrected multi-layer stochastic blockmodel}

Our next set of quality functions are based on a statistical model of random multi-layer networks which we call the degree corrected multi-layer stochastic blockmodel (DCMLSBM). This model can be thought of as a model with community structure that is built upon the multi-layer expected degree models introduced in Section 2 as null models. Both the multi-layer stochastic blockmodel (MLSBM) and the DCMLSBM have been previously used in the literature as a statistical model for multi-layer networks with block structures \citep{valles14, hxa14,peixoto15,pc15,taylor15,stanley15}. In what follows we first define various degree corrected extensions of the two models analyzed in \citet{pc15}, the MLSBM and the restricted MLSBM (RMLSBM) based on the multi-layer expected degree null models defined earlier, and then derive likelihood modularity functions \citep{bc09,kn11} from them. Several generative models based on multi-layer extensions of SBM were developed in \citet{peixoto15} with priors on the parameters and a Bayesian model selection procedure was developed. In this paper we restrict ourselves to variations only in terms of degree through the SD and ID multi-layer expected degree null models described in Section 2 and variation in terms of block parameters through a restriction similar to \citet{pc15}.

It has been argued previously in the literature that the modularities based on the single layer SBM and DCBM are more widely applicable than ad hoc forms of quality functions and often remedy some of the deficiencies of the later \citep{ball11, bc09}. Since the Newman-Girvan modularities consider only the intra-community edges and do not take into account the inter-community edge structure directly (although they are used indirectly to compute the expected intra-community edges), they miss some of the information taken into account by the likelihood modularities which consider both intra and inter community edges. As an example, the NG modularities fail to detect dissortative mixing/heterophilic communities and perform poorly if the community sizes are unbalanced, while likelihood modularities are robust to such cases.

Similar to the single layer stochastic blockmodel, the multi-layer stochastic blockmodel also assumes stochastic equivalence of nodes for a given type of edge within each community and hence fails to model real life multi-layer networks with degree heterogeneity. To remedy the situation for single layer graphs, degree corrected blockmodel (DCBM) was proposed by \citet{kn11}. They also characterized the modularity based on this model as a Kullback-Leibler divergence between this model and a null model without the community structure. Such a null model would be equivalent to the Chung-Lu expected degree random graph model. Hence both the MLSBM and RMLSBM should be corrected for degree variation using multi-layer extensions of the expected degree null models described earlier in (\ref{IDexpected}) and (\ref{SDexpected}).

Throughout the section we assume that the edges $A_{ij}^{(m)}$ between two nodes $i$ and $j$ are formed independently following Poisson distribution, given the community assignments of the nodes $z_{i}$ and $z_{j}$ and the degree vectors of the two nodes $\mathbf{k}_{i}$ and $\mathbf{k}_{j}$:
\[
A_{ij}^{(m)}|(z_{i}=q,z_{j}=l)\sim Poisson(P_{ij}^{(m)}).
\]
We model the Poisson mean parameter for the multi-layer stochastic blockmodel in four different ways with varying number of parameters. The first two of the models have the independent degree (ID) expected degree model as their null model.
The first model is parameterized by node-layer specific parameters $\theta_i^{(m)}$ similar to the ID model and community-layer specific parameters $\pi_{ql}^{(m)}$. Formally,
 \begin{equation}
 P_{ij}^{(m)}=\theta^{(m)}_{i}\theta^{(m)}_{j}\pi_{ql}^{(m)},\quad \quad  i,j\in\{1,\ldots,N\},\: m\in\{1,\ldots,M\}, \; q,l\in\{1,\ldots,K\},
\end{equation}
with the identifiability constraints
 \begin{equation*}
\sum_{i: z_{i}= q} \theta^{(m)}_{i}  =1, \quad  m\in\{1,\ldots,M\}, \: q \in\{1,\ldots,K\}.
 \end{equation*}
We call this model the degree corrected multi-layer stochastic blockmodel (DC-MLSBM). 

The next model is the degree corrected version of the RMLSBM, which we call the DC-RMLSBM,
 \begin{equation}
  P_{ij}^{(m)}=\theta^{(m)}_{i}\theta^{(m)}_{j}\pi_{ql}, \quad \quad   i,j\in\{1,\ldots,N\}, \: m\in\{1,\ldots,M\}, \: q,l\in\{1,\ldots,K\},
\end{equation}
with identifiability constraints
 \begin{equation*}
 \sum_{i: z_i = q} \theta^{(m)}_{i} =1, \quad m\in\{1,\ldots,M\}, \: q \in\{1,\ldots,K\}.
 \end{equation*}
This model has node-layer specific parameter $\theta_i^{(m)}$ similar to DC-MLSBM, while the community specific parameters $\pi_{ql}$ are the same across layers. The DC-MLSBM and DC-RMLSBM have $M(N-K)+MK(K+1)/2$ and $M(N-K)+K(K+1)/2$ parameters respectively.

In the next two models the underlying null model is the shared degree (SD) expected degree model, and hence the node specific ``degree" parameter $\theta_i$ is common across the layers. We call the models the shared degree multi-layer stochastic blockmodel (SD-MLSBM) and the shared degree restricted multi-layer stochastic blockmodel (SD-RMLSBM) respectively. The SD-MLSBM is
 \begin{equation}
P_{ij}^{(m)}=\theta_{i}\theta_{j} \pi^{(m)}_{ql}, \quad \quad  i,j\in\{1,\ldots,N\}, \: m\in\{1,\ldots,M\}, \: q,l\in\{1,\ldots,K\},
\end{equation}
with identifiability constraints
 \begin{equation*}
 \sum_{i: z_i = q} \theta_{i}=1, \quad   q \in\{1,\ldots,K\},
 \end{equation*}
and the SD-RMLSBM is
 \begin{equation}
P_{ij}^{(m)}=\theta_{i}\theta_{j} \beta_{m} \pi_{ql}, \quad \quad  i,j\in\{1,\ldots,N\}, \: m\in\{1,\ldots,M\}, \: q,l\in\{1,\ldots,K\},
\end{equation}
with identifiability constraints
 \begin{align*}
\sum_{m} \beta_{m} =1, \quad \quad \sum_{i: z_i = q} \theta_{i}=1, \quad  q \in\{1,\ldots,K\}.
 \end{align*}
The SD-MLSBM and SD-RMSBM have $N-K+MK(K+1)/2$ and $N-K+M-1+K(K+1)/2$ parameters respectively. 

Clearly the four models are nested models with different number of parameters. We consider an asymptotic scenario here to estimate the number of parameters each of the models will have asymptotically. We consider asymptotics as both $M$ and $N$ grow. However, we do not assume any relationship between them, nor do we require both of the them to grow simultaneously. For example, it could be the case that only $N$ grows or only $M$ grows. Let the growth rate of communities be $K=O(f(N),\ g(M))$ for some functions $f$ and $g$ with the constraint that $K \leq N$, since the number of communities cannot be larger than the number of nodes.
\begin{itemize}
    \item Let $f(N)=1$, $g(M)=1$, i.e., $K=O(1)$, which implies that $K$ does not grow with $N$ or $M$. Then DC-MLSBM, DC-RMLSBM have  $O(MN)$ parameters while SD-MLSBM and SD-RMLSBM have $O(N+M)$ parameters.
    \item Let $f(N)=N^{1/2}$, $g(M)=1$, i.e., $K=O(N^{1/2})$, which implies that $K$ grows with $N$ but does not grow with $M$. Then DC-MLSBM, DC-RMLSBM and SD-MLSBM have  $O(MN)$ parameters while SD-RMLSBM has $O(N+M)$ parameters.
    \item Let $f(N)=1$, $g(M)=M^{1/2}$, i.e., $K=O(M^{1/2})$, which implies that $K$ grows with $M$ but does not grow with $N$. However, here we need to assume that $M^{1/2} <N$ to make sure that $K<N$ holds. Then DC-MLSBM, DC-RMLSBM have  $O(MN + M^2)$ parameters, SD-MLSBM has $O(M^2+N)$ parameters, while SD-RMLSBM has $O(M+N)$ parameters.
\end{itemize}

From the preceding discussion we see that SD-RMLSBM has fewer parameters in most asymptotic scenarios and in some case by order of magnitude. In general, for the models that have considerably fewer number of parameters (e.g., SD-RMLSBM), we expect the maximum likelihood estimates to have less variance at the expense of some bias. Such gain in terms of low variance at the expense of bias would be advantageous in situations where the network layers are sparse \citep{pc15}. On the other hand, when network layers are dense, we expect SD-RMLSBM to underperform compared to a model (e.g.  DC-MLSBM) which has less bias.

\subsection{Likelihood quality functions}

To derive quality functions based on the four models defined above, we take the profile likelihood approach similar to \citet{bc09} and \citet{kn11}, where we maximize the conditional log-likelihood $l(A|z;P)$ of the adjacency matrix given the group assignments. This is done by plugging in the MLE of the parameter set $P$ conditional on $z$.  The conditional log-likelihood for DC-MLSBM can be written as (dropping the terms that do not depend on the class assignment)
\begin{align}
l(A;z,\pi, \theta)&=\frac{1}{2}\sum_{m=1}^{M}\sum_{i,j}\{A_{ij}^{(m)} \{\log (\pi_{z_{i}z_{j}}^{(m)})+\log(\theta^{(m)}_{i})+\log(\theta^{(m)}_{j})\} -\theta^{(m)}_{i}\theta^{(m)}_{j} \pi_{z_{i}z_{j}}^{(m)} \} \nonumber \\
&= \sum_{m} \sum_{i} k_{i}^{(m)} \log (\theta_{i}^{(m)}) + \sum_{m} \sum_{q \leq l} \{ e_{ql}^{(m)} \log (\pi^{(m)}_{ql})  - \pi^{(m)}_{ql} \}.
\label{eq:DCMLSBMllk}
\end{align}
The MLE of $\pi$ can be obtained by a straightforward differentiation of the log-likelihood function. However to find the MLE of $\theta$ under the identifiability constraints, we need to use the Lagrange multipliers as follows. The objective function to be optimized is
\begin{equation*}
\Lambda(\theta,\lambda)=\sum_{i} \sum_{m} k_{i}^{(m)} \log (\theta_{i}^{(m)}) + \sum_{m} \sum_{q} \lambda_{mq} (\sum_{i: z_i = q} \theta^{(m)}_{i} -1).
\end{equation*}
Solving for $\theta$ and $\lambda$ we obtain the following solutions for the MLE:
\begin{align*}
\hat{\theta}^{(m)}_{i}=\frac{k_{i}^{(m)}}{\sum_{i: z_i = q} k_{i}^{(m)}}=\frac{k_{i}^{(m)}}{e_{q}^{(m)}}, \quad \quad \quad
 \hat{\pi}_{ql}^{(m)}=\sum_{i,j :\ z_i = q, z_j = l} A_{ij}^{(m)}= e_{ql}^{(m)}.
\end{align*}
Plugging in these estimates into the log-likelihood function gives the maximized log-likelihood function as
\begin{align}
l(A;z)&=\sum_{i} \sum_{m} k_{i}^{(m)} \log \left( \frac{k_{i}^{(m)}}{e_{q}^{(m)}} \right)  + \sum_{m} \sum_{q \leq l} \{ e_{ql}^{(m)} \log (e_{ql}^{(m)})  - e_{ql}^{(m)} \}  \\
&=\sum_{m} \sum_{q \leq l}  e_{ql}^{(m)} \log (e_{ql}^{(m)}) -\sum_{m} \sum_{q \leq l} e_{ql}^{(m)} + \sum_{i} \sum_{m} k_{i}^{(m)} \log (k_{i}^{(m)})-\sum_{q} \sum_{m} e_{q}^{(m)} \log (e_{q}^{(m)}) \nonumber.
\end{align}
Now ignoring the terms that do not depend on the class assignment (the 2nd and 3rd terms), we get
\begin{equation}
l(A;z)= \sum_{m} \sum_{q \leq l} e_{ql}^{(m)} \log (e_{ql}^{(m)}) -\sum_{q} \sum_{m} e_{q}^{(m)} \log (e_{q}^{(m)}).
\end{equation}
It is easy to see that this maximized likelihood function can be written as
\begin{equation*}
 Q=\sum_{m} \sum_{q \leq l} \Bigg \{e_{ql}^{(m)} \log \left(\frac{e_{ql}^{(m)}}{e_{q}^{(m)} e_{l}^{(m)}} \right) \Bigg \},
\end{equation*}
which we call the un-normalized likelihood quality function.

However, the quality function at this form take more contribution from denser layers as compared to the sparser ones and are not appropriate for community detection in multi-layer networks. Since we are interested in inference about the underlying community structure across the layers, we want to capture the ``signals" available from each layer irrespective of its density and combine them together. Hence we need to normalize this likelihood quality function layer-wise. The role of normalization is especially important in situations where the layers of a network represent quite different relationships. In those situations it may happen that a dense network is uninformative and a sparse one is quite informative. The deficiencies in un-normalized likelihood quality are conceptually the same as those for which an aggregate of adjacency matrices across layers fails to detect the community signal. Apart from reducing the undue influence of highly dense layers on the objective function, normalization helps to retain the Kullback-Liebler (KL) divergence based probabilistic interpretation of likelihood quality \citep{kn11}. Since by assumption, given the label assignments of the nodes, the network layers are formed independently each with a Poisson distribution, the KL divergence of the model with block structure from a null model without a block structure is an indicator of the goodness of fit for that model in each of the network layers. Hence much like the configuration model case, the multi-layer likelihood quality function in the stochastic blockmodel case can also be viewed as an aggregation of divergences in the component networks. For this purpose we normalize $A^{(m)}_{ij}$ by twice the total number of edges in the $m$th layer, $2L^{(m)}$. Consequently quantities that are derived from $A$ also gets normalized accordingly.
The likelihood quality function after proper normalization can be written as
\begin{equation}
Q_{DC-MLSBM} =\sum_{m} \sum_{q \leq l} \Bigg \{\frac{e_{ql}^{(m)}}{2L^{(m)}} \log \left(\frac{e_{ql}^{(m)}/2L^{(m)}}{(e_{q}^{(m)}/2L^{(m)}) (e_{l}^{(m)}/2L^{(m)})} \right) \Bigg \}. \label{QDCMLSBM}
\end{equation}

Similarly for DC-RMLSBM, the conditional likelihood along with the constraints can be simplified as (dropping the terms not dependent on the parameters)
\begin{align}
l(A;z,\pi, \theta)&=\frac{1}{2}\sum_{m=1}^{M}\sum_{i,j}\{A_{ij}^{(m)} \{\log (\pi_{z_{i}z_{j}}) +\log(\theta^{(m)}_{i})+\log(\theta^{(m)}_{j})\}-\theta^{(m)}_{i}\theta^{(m)}_{j} \pi_{z_{i}z_{j}} \}  \nonumber \\
&= \sum_{i} \sum_{m} k_{i}^{(m)} \log (\theta_{i}^{(m)}) + \sum_{m} \sum_{q \leq l} \{ e_{ql}^{(m)} \log (\pi_{ql})  - \pi_{ql} \}. \label{eq:DCRMLSBMllk}
\end{align}
The MLEs of $\theta$ and $\pi$ under the constraints are once again obtained by the method of Lagrange multipliers as explained before:
\begin{align*}
\hat{\theta}^{(m)}_{i}=\frac{k_{i}^{(m)}}{\sum_{i: z_i= q} k_{i}^{(m)}}=\frac{k_{i}^{(m)}}{e_{q}^{(m)}}, \quad \quad \quad
 \hat{\pi}_{ql}=\sum_{m} \sum_{i,j :\ z_i= q, z_j = l} A_{ij}^{(m)}= \sum_{m} e_{ql}^{(m)}.
\end{align*}
The profile likelihood quality function function can be obtained by plugging in the MLEs into the log-likelihood equation and then dropping the terms that do not depend on the class assignment:
\begin{equation}
l(A;z)= \sum_{q \leq l} \sum_{m}  e_{ql}^{(m)} \log (\sum_{m} e_{ql}^{(m)}) -\sum_{q} \sum_{m} k_{q}^{(m)} \log (e_{q}^{(m)}),
\end{equation}
and with proper normalization the likelihood quality function is
\begin{equation}
Q_{DC-RMLSBM} =\sum_{m} \sum_{q \leq l} \Bigg \{\frac{e_{ql}^{(m)}}{2L^{(m)}} \log \left(\frac{\sum_{m} (e_{ql}^{(m)}/2L^{(m)})}{(e_{q}^{(m)}/2L^{(m)}) (e_{l}^{(m)}/2L^{(m)}) } \right)\Bigg \}. \label{QDCRMLSBM}
\end{equation}

The likelihood quality functions for the two shared degree models can also be derived in a similar fashion.  For SD-MLSBM, the conditional log-likelihood without the terms independent of the parameters is
\begin{align}
l(A;z,\pi, \theta)&=\frac{1}{2}\sum_{m=1}^{M}\sum_{i,j}\{A_{ij}^{(m)} \{\log (\pi^{(m)}_{z_{i}z_{j}}) +\log(\theta_{i})+\log(\theta_{j})\} -\theta_{i}\theta_{j} \pi^{(m)}_{z_{i}z_{j}} \} \\ \nonumber
&= \sum_{i} \sum_{m} k_{i}^{(m)} \log (\theta_{i}) + \sum_{m} \sum_{q \leq l} \{e_{ql}^{(m)} \log ( \pi^{(m)}_{ql} ) - \pi^{(m)}_{ql} \}. \label{eq:M3llk}
\end{align}
The maximum likelihood estimates of the parameters are
\begin{align*}
\hat{\theta}_{i}  =\frac{\sum_{m}k_{i}^{(m)}}{\sum_{m}\sum_{i: z_i= q} k_{i}^{(m)}}=\frac{\sum_{m}k_{i}^{(m)}}{\sum_{m} e_{q}^{(m)}}, \quad \quad \quad
 \hat{\pi}^{(m)}_{ql}  = \sum_{i,j :\ z_i= q, z_j= l} A_{ij}^{(m)}= e_{ql}^{(m)},
\end{align*}
and hence the profile likelihood  quality function with normalization is
\begin{equation}
Q_{SD-MLSBM} =\sum_{m} \sum_{q \leq l} \Bigg \{\frac{e_{ql}^{(m)}}{2L^{(m)}} \log \left(\frac{ (e_{ql}^{(m)}/2L^{(m)})}{\sum_{m} (e_{q}^{(m)}/2L^{(m)}) \sum_{m} (e_{l}^{(m)}/2L^{(m)})} \right)\Bigg \}. \label{SDMLSBM}
\end{equation}

For SD-RMLSBM, the conditional log-likelihood without the terms independent of the parameters is
\begin{align}
l(A;z,\pi, \theta)&=\frac{1}{2}\sum_{m=1}^{M}\sum_{i,j}\{A_{ij}^{(m)} \{\log (\pi_{z_{i}z_{j}}) +\log(\beta_{m})+\log(\theta_{i})+\log(\theta_{j})\} -\theta_{i}\theta_{j} \beta_{m} \pi_{z_{i}z_{j}} \} \\ 
&= \sum_{i} \sum_{m} k_{i}^{(m)} \log (\theta_{i}) + \sum_{m} \sum_{q \leq l} e_{ql}^{(m)} \{ \log (\beta_{m}) +\log (\pi_{ql})\} -\sum_{q\leq l} \pi_{ql} \nonumber. \label{eq:M4llk}
\end{align}
The maximum likelihood estimates of the parameters are
\begin{align*}
\hat{\theta}_{i}  =\frac{\sum_{m}k_{i}^{(m)}}{\sum_{m}\sum_{i : z_i= q} k_{i}^{(m)}} & =\frac{\sum_{m}k_{i}^{(m)}}{\sum_{m}e_{q}^{(m)}}, \quad \quad  \hat{\beta}_{m} =\frac{\sum_{q \leq l} e_{ql}^{(m)}}{\sum_{m} \sum_{q\leq l} e_{ql}^{(m)}}=\frac{L^{(m)}}{L}, \\
 \hat{\pi}_{ql} & =\sum_{m} \sum_{i,j:\ z_i = q, z_j = l} A_{ij}^{(m)}= \sum_{m} e_{ql}^{(m)}.
\end{align*}
Ignoring the terms not dependent on the label assignments and after normalization, the likelihood quality function is
\begin{equation}
Q_{SD-RMLSBM} =\sum_{m} \sum_{q \leq l} \Bigg \{ \frac{e_{ql}^{(m)}}{2L^{(m)}} \log \left(\frac{\sum_{m} (e_{ql}^{(m)}/2L^{(m)})}{\sum_{m} (e_{q}^{(m)}/2L^{(m)}) \sum_{m} (e_{l}^{(m)}/2L^{(m)}) } \right)\Bigg \}. \label{SDRMLSBM}
\end{equation}

\section{Computation}
We adapt the Louvain algorithm \citep{blondel08} to multi-layer network settings for computing both the number of communities and the optimal partitions using the multi-layer configuration model (MLCM) based multi-layer modularities from Section 3. Similar to the original Louvain algorithm, the modified algorithm is also a two-phased fast greedy optimization method for community detection.  For optimizing the likelihood quality function based measures from Section 4, we implement a multi-layer version of the algorithm used by \citet{kn11}. As in \cite{kn11}, we need the number of communities $K$ to be known in advance for this method. The algorithm is a Kernighan-Lin type graph partitioning algorithm and is a non-greedy approach which leads to more accurate results for a known $K$ than the Louvain approach. It however requires a starting partition and the final solution depends on the quality of the initial value. This algorithm often gets stuck in a local minimum and hence we use multiple starting points to improve the quality of partitions. 

We refer the reader to \citet{blondel08} and \citet{kn11} for details of the algorithms. For both algorithms the execution speed heavily depends on the ability to quickly compute the increase in modularity score for a one step change without having to re-compute the modularity value for the entire network. For the Louvain algorithm, this one step change is the increase in modularity for removing a node $i$ from its own community (i.e., the community which only contains $i$) and moving it to the community of one of its neighbors $j$, say community $q$.  For each one of the modularities, we have derived this one step change. Here we only give example formulas for three MLCM based modularities to compute the one step change in Louvain algorithm. We define an additional notation. Let $d^{(m)}_{iq}$ denote the number of type $m$ edges from node $i$ to a neighboring community $q$. Then
\begin{align*}
\Delta Q_{MNavrg} &= \sum_{m} \frac{1}{2L^{(m)}} \Bigg \{d^{(m)}_{i,q}-\frac{e^{(m)}_{q}k^{(m)}_{i}}{L^{(m)}} \Bigg \}, \nonumber \\
\Delta Q_{SDavrg} &= \sum_{m} \frac{1}{2L^{(m)}}\Bigg \{d^{(m)}_{i,q}-\frac{L^{(m)}}{L}\frac{(\sum_{m} e^{(m)}_{q}) (\sum_{m} k^{(m)}_{i})}{L} \Bigg \}, \nonumber \\
\Delta Q_{SDlocal} &= \sum_{m} \frac{1}{2L^{(m)}}\Bigg \{d^{(m)}_{i,q}-\frac{k^{(m)}_{i} (\sum_{m} e^{(m)}_{q}) + (\sum_{m} k^{(m)}_{i}) e^{(m)}_{q} }{2L} \Bigg \}.
\end{align*}

The Kernighan-Lin type algorithm for optimizing the likelihood quality functions starts with an initial assignment of nodes to the $K$ communities, and loops through each node to consider moving it from its current community to any of the other $K-1$ candidate communities \citep{kn11}. A move is made if the likelihood is increased as a result of that move. Once all nodes have been considered (each node is checked only once), we arrive at a new community assignment and repeat this process until the likelihood does not increase beyond a pre-specified tolerance.  The key to this algorithm is fast computation of change in likelihood due to moving node $i$ from its community $r$ to another neighboring community $s$ \citep{kn11}. In particular, we do not need to compute the full likelihood value at each iteration because the difference can be found as a function involving only a few quantities. Here we provide an expression for the change for DC-MLSBM, which serves as an example. Define the function $f(x) =x \log x$ . Then 
\begin{align*}
    \Delta L & = \Delta L_{rem} + \Delta L_{to} + \sum_m 2 \left\{f\left(\frac{e_{rs}^{(m)} + d_{i,r}^{(m)} - d_{is}^{(m)}}{2L^{(m)}}\right) - f\left(\frac{e_{rs}^{(m)}}{2L^{(m)}}\right)\right\},\\
    \Delta L_{rem} & = \sum_m \Bigg \{\sum_{t \neq r,s}2\left\{f\left(\frac{e_{rt}^{(m)}  - d_{it}^{(m)}}{2L^{(m)}}\right) - f\left(\frac{e_{rt}^{(m)}}{2L^{(m)}}\right)\right\} \\
    & \quad + f\left(\frac{e_{rr}^{(m)}  - (d_{ir}^{(m)} + A_{ii}^{(m)})}{L^{(m)}}\right) - f\left(\frac{e_{rr}^{(m)} }{L^{(m)}}\right) - 2\left\{f\left(\frac{e_{r}^{(m)}  - k_{i}^{(m)}}{2L^{(m)}}\right) - f\left(\frac{e_{r}^{(m)}  }{2L^{(m)}}\right)\right\}
    \Bigg \}, \\
        \Delta L_{to} & = \sum_m \Bigg\{\sum_{t \neq r,s}2\left\{f\left(\frac{e_{st}^{(m)}  + d_{it}^{(m)}}{2L^{(m)}}\right) - f\left(\frac{e_{st}^{(m)}}{2L^{(m)}}\right)\right\} \\
    & \quad + f\left(\frac{e_{ss}^{(m)}  + (d_{is}^{(m)} + A_{ii}^{(m)})}{L^{(m)}}\right) - f\left(\frac{e_{ss}^{(m)}  }{L^{(m)}}\right) - 2\left\{f\left(\frac{e_{s}^{(m)}  + k_{i}^{(m)}}{2L^{(m)}}\right) - f\left(\frac{e_{s}^{(m)}  }{2L^{(m)}}\right)\right\}
    \Bigg \}.
\end{align*}
We have implemented these methods in Python and the codes are available at \url{https://u.osu.edu/subhadeep/codes/}.

\section{Simulation results}
In this section we numerically compare the performance of the various  quality functions for community detection through a simulation study.
Since the true class labels of the nodes are known in simulated data,
we compare the class assignments from different methods with the true
labels. This comparison involves two stages. Since the Louvain algorithm applied to the modularities can identify the number of clusters automatically, an effective community detection in situations where the number of communities is unknown must first identify the number of communities correctly. Hence the first step of comparison is in terms of the number of communities detected. The metric used for this purpose is the mean square error of the number of classes recovered across the repetitions. The second step would be to compare the goodness of the class assignments. As a metric, we use the normalized mutual information (NMI) which is an information theory based similarity measure between two cluster assignments \citep{kn11}. This metric takes values between $0$ and $1$, where, in theory $0$ indicates that the class assignment is random with respect to the true class labels and $1$ indicates a perfect match with the true class labels. However, we note that the NMI score can take a small positive value even when the estimated cluster assignment is random with respect to the true cluster assignment purely due to chance. Since the measure is ``normalized" it can be used to compare clustering solutions with different number of clusters as well. Finally assuming that the number of clusters is known in advance, we compare the clustering accuracy of the modularity scores  in terms of NMI. All the results reported throughout the section are the average of the metric across $40$ repetitions of the simulations.

We then compare the relative performance of the multi-layer modularities along with a baseline method for comparison, the NG modularity on the aggregated adjacency matrix. The comparison is performed under
various settings on the number of nodes $N$, the number of communities $K$ and the average degree per layer.

\subsection{Data generation}
We generate data from both the multi-layer stochastic blockmodel and its degree corrected version. For this purpose, we first generate $N$ node labels independently from a $K$ class multinomial distribution. The network community sizes  are varied by varying the parameters of the multinomial distribution with equal parameters leading to ``balanced" communities.  We next generate the $M$ layers using the stochastic blockmodel each with a different connectivity matrix. In our stochastic blockmodel, the connectivity matrices give
larger probabilities for intra-block edges in comparison
to the inter-block edges.  The general structure of the connectivity matrix has $(\rho\lambda_{1}, \ldots , \rho\lambda_{k} )$  in the diagonal and the same element $\rho \epsilon$ in the off diagonal. We control the signal strength in different layers by varying the ratio of $\lambda$'s with $\epsilon$ from layer to layer while we control the average degree per layer by varying the parameter $\rho$. Throughout the section strong signal means that each of $\lambda_{i}$ is roughly 3-4 times greater than $\epsilon$ and weak signal means each of  $\lambda_{i}$ is only marginally greater than $\epsilon$. If the degree corrected stochastic blockmodel is used for data generation, then the degree parameter is generated using a power law distribution, one parameter for each node in the shared degree model and one parameter for each node in each layer in the independent degree model.

\subsection{Number of communities unknown}

In our first simulation we assume the number of communities is unknown and use the Louvain algorithm to automatically determine the number of communities along with the partition. We consider two scenarios in terms of the composition of the component layers: the first one having sparse strong signal in all layers and the second one having mixed sparsity and signal quality in its layers where strong and mixed are as described in the previous paragraph.

\subsubsection{Sparse strong signal}

\begin{figure}[!h]
\begin{subfigure}{0.5\textwidth}
\centering{}
\includegraphics[width=.95\linewidth]{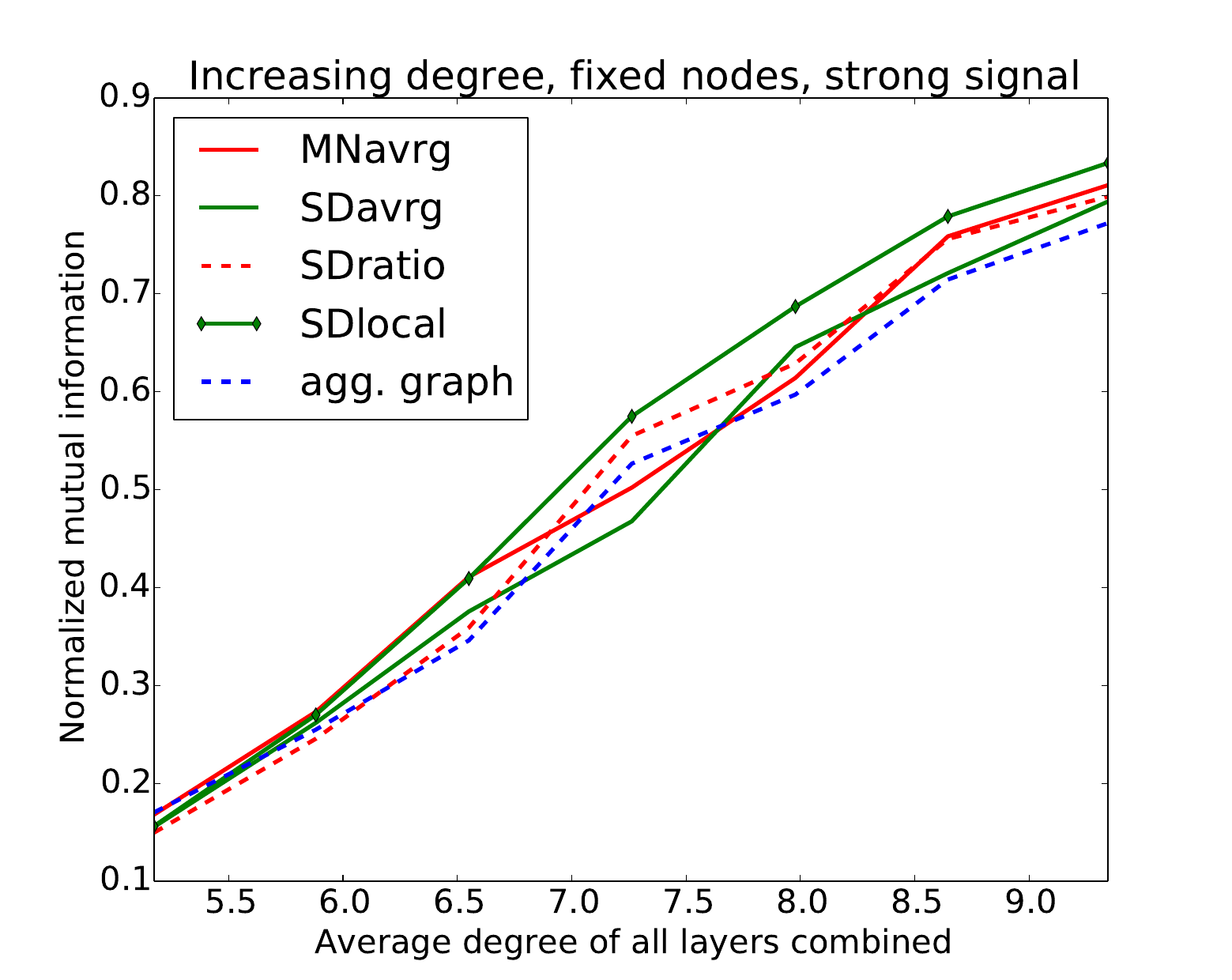}
\end{subfigure}%
\begin{subfigure}{0.5\textwidth}
\centering{}
\includegraphics[width=.98\linewidth]{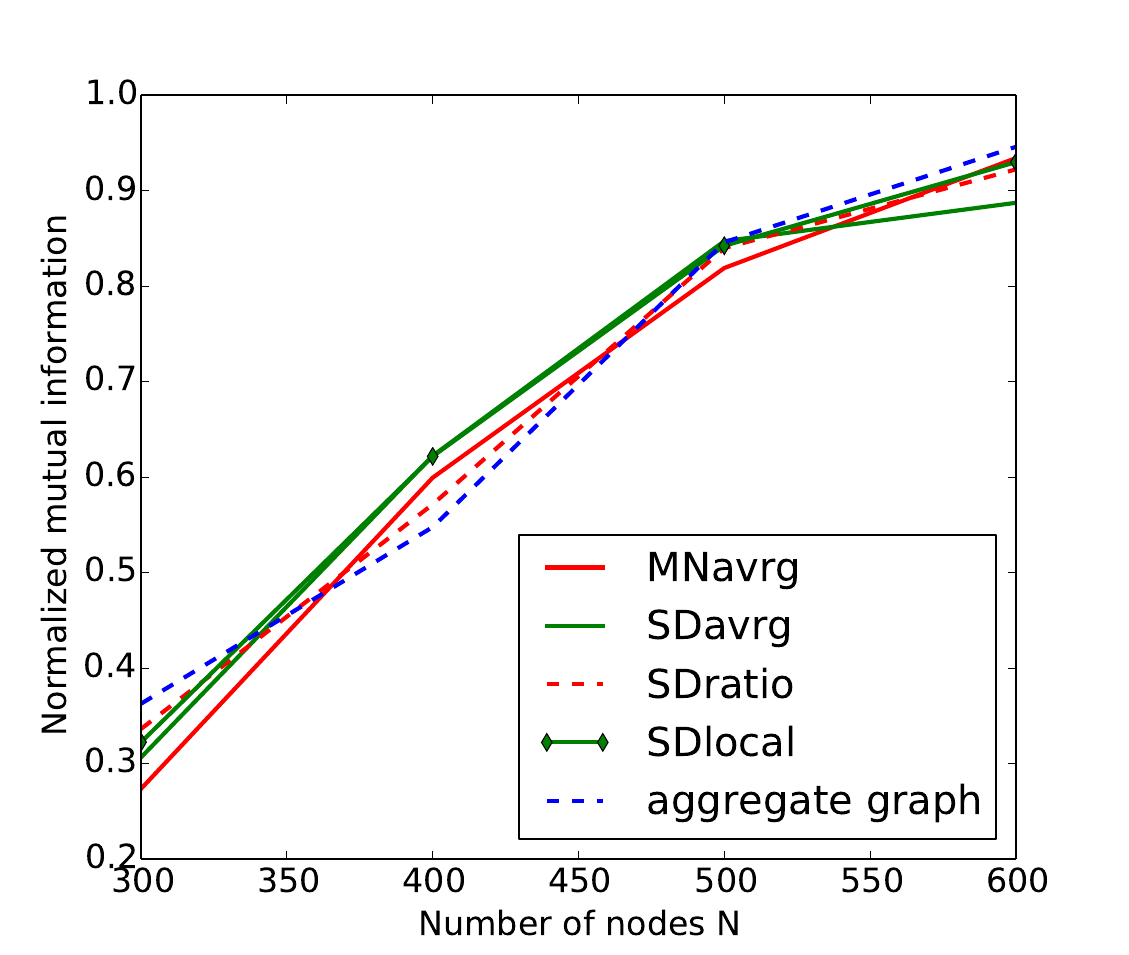}
\end{subfigure}
\begin{subfigure}{0.5\textwidth}
\centering{}
\includegraphics[width=.95\linewidth]{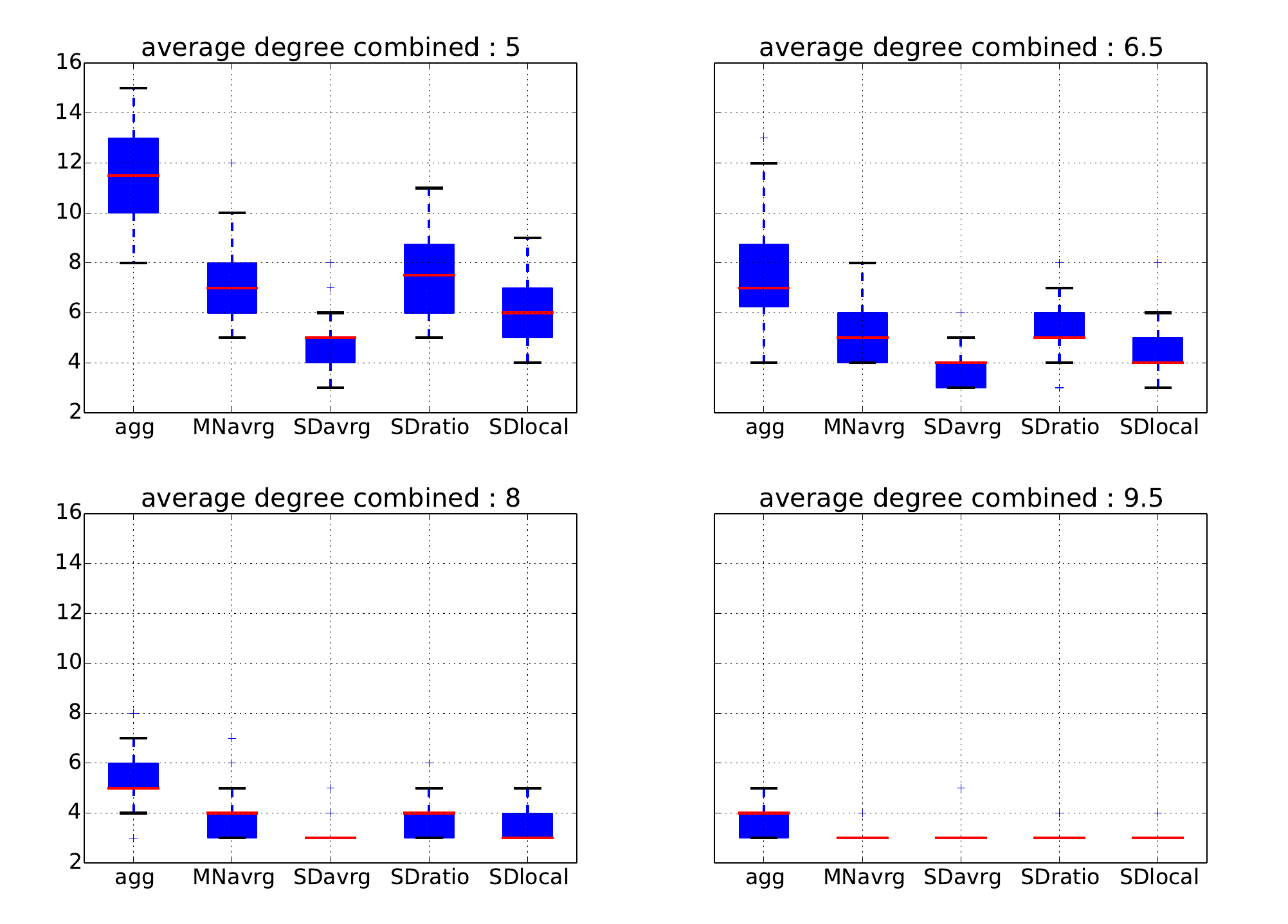}
\end{subfigure}%
\begin{subfigure}{0.5\textwidth}
\centering{}
\includegraphics[width=.98\linewidth]{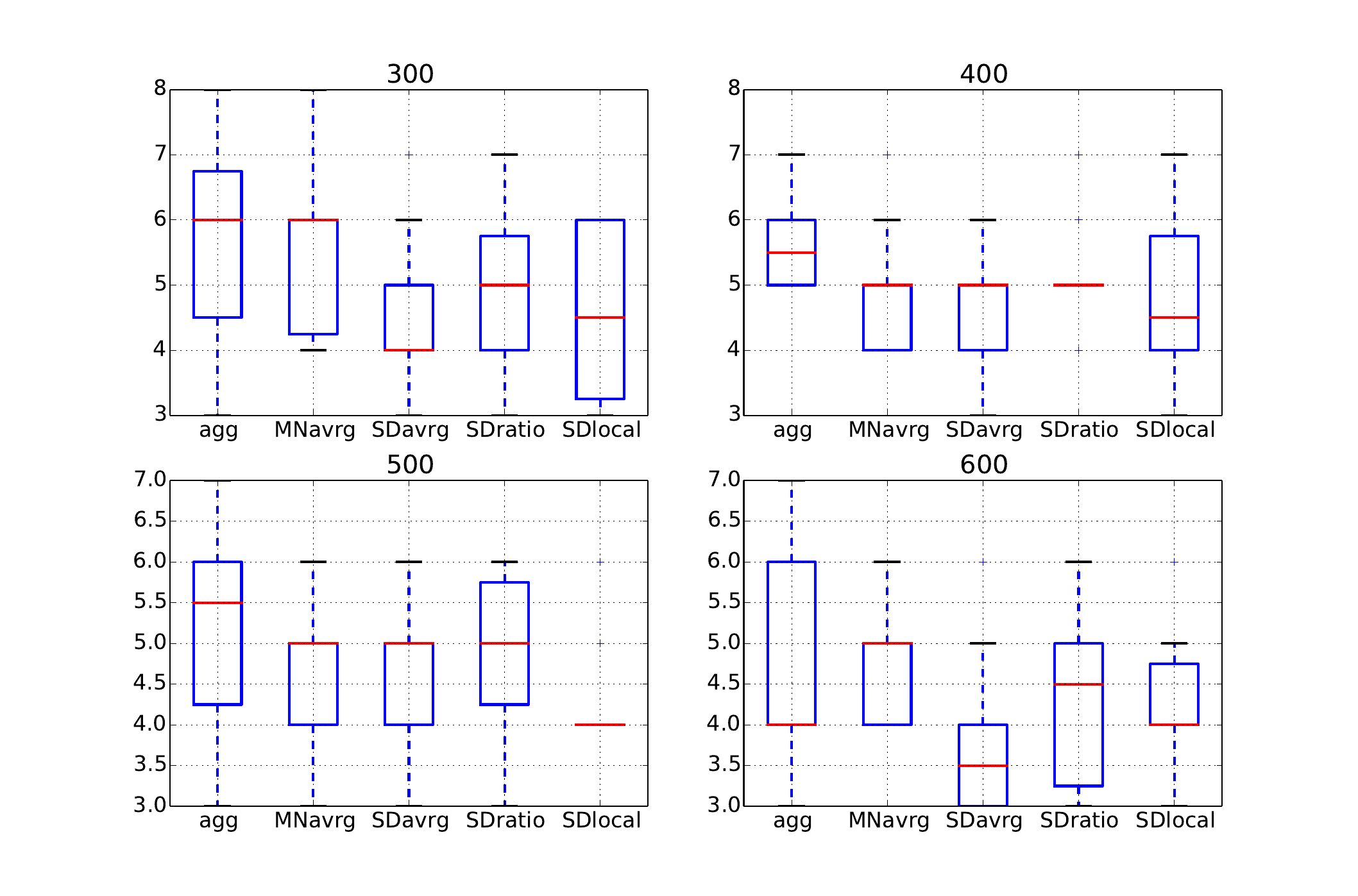}
\end{subfigure}
\vspace{-10pt}
\begin{center} (a) \hspace{200pt} (b) \end{center}
\vspace{-10pt}
\caption{Comparison of the performance of various multi-layer modularities for data generated from the DCBM with independent degrees. The layers are sparse and the signal is strong across all layers. (a) The number of nodes and number of communities are fixed at 800 and 3 while the average degree of the nodes across all layers combined is increased. (b) The number of nodes is increased from 300 to 600 while the number of communities is fixed at 3. In both cases, the top figure is the comparison in terms of NMI of the community assignment and the bottom figure is the box plot of the number of communities detected.}
\label{sparse}
\end{figure}

With all component layers being sparse and strong in signal quality, we design two simulation scenarios. First we fix the number of communities $K$ at 3 and the number of nodes $N$ at 800 while we vary the average density of the multi-layer graph. Figure \ref{sparse}(a) shows the results of this simulation. The top figure is a comparison in terms of NMI of the community assignment and the bottom figure is a box plot of the number of communities detected. While there is not much difference among the modularities compared in terms of the NMI, the box plots for the number of classes detected show that the shared degree methods SDavrg and SDlocal are closer to the correct number of communities (which in this case is 3) as compared to the MNavrg and the aggregate graph in sparser networks. SDavrg and SDlocal also converge to the correct number of communities faster than MNavrg and aggregate graph as the network becomes denser.

In the second simulation scenario, we fix the number of communities at $3$ and vary the number of nodes from $300$ to $600$. Figure \ref{sparse}(b) shows the results of this simulation where as before the top figure is the comparison in terms of NMI of the community assignment and the bottom figure is the box plot of the number of communities detected. Similar to the previous case, we observe that the number of communities detected by SDavrg and SDlocal converges to the true number of communities faster than MNavrg and aggregate graph as the number of nodes increases.

\subsubsection{Mixed signal layers}

\begin{figure}[!ht]
\begin{subfigure}{0.48\textwidth}
\centering{}
\includegraphics[width=.95\linewidth]{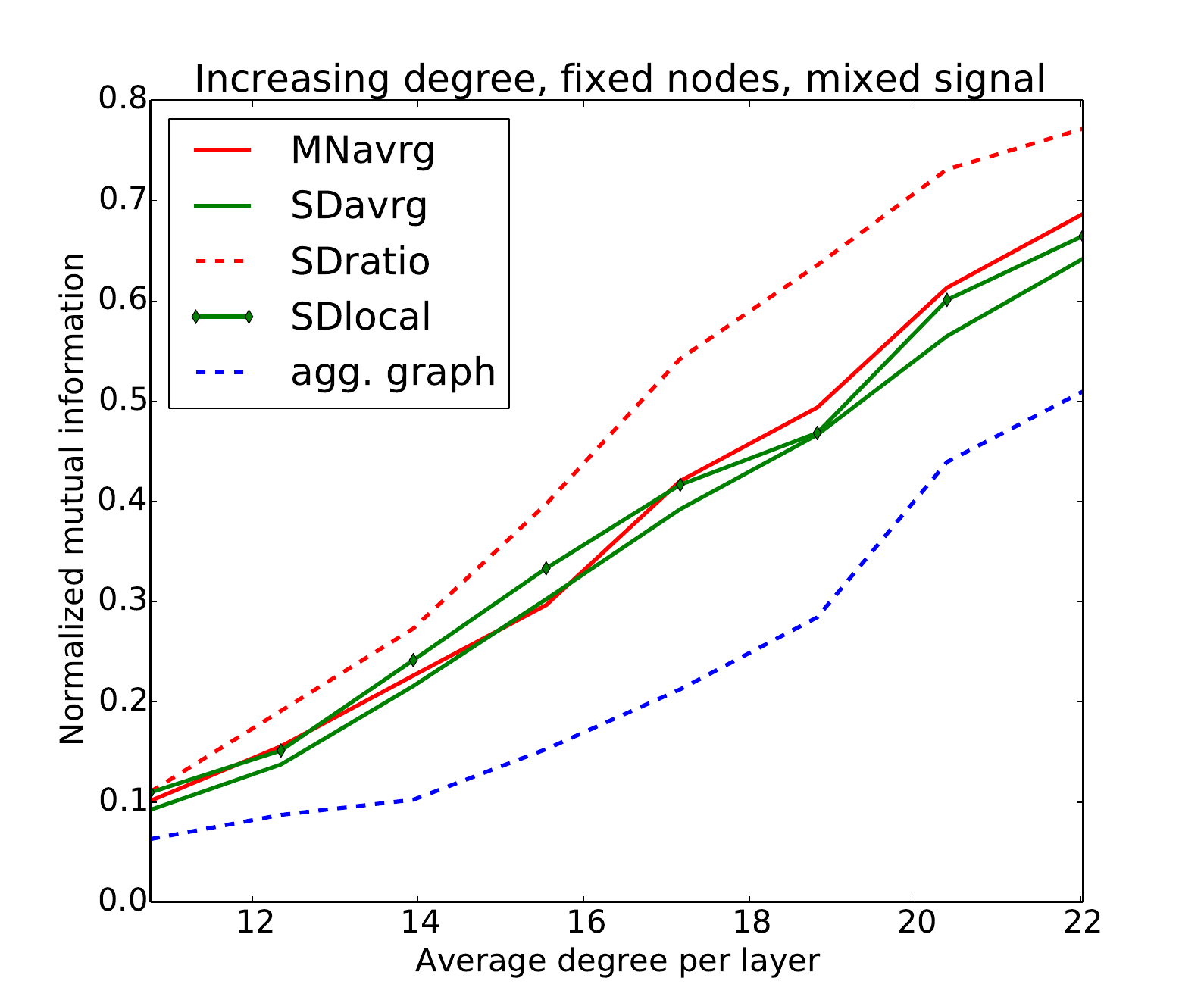}
\end{subfigure}%
\begin{subfigure}{0.48\textwidth}
\centering{}
\includegraphics[width=.95\linewidth]{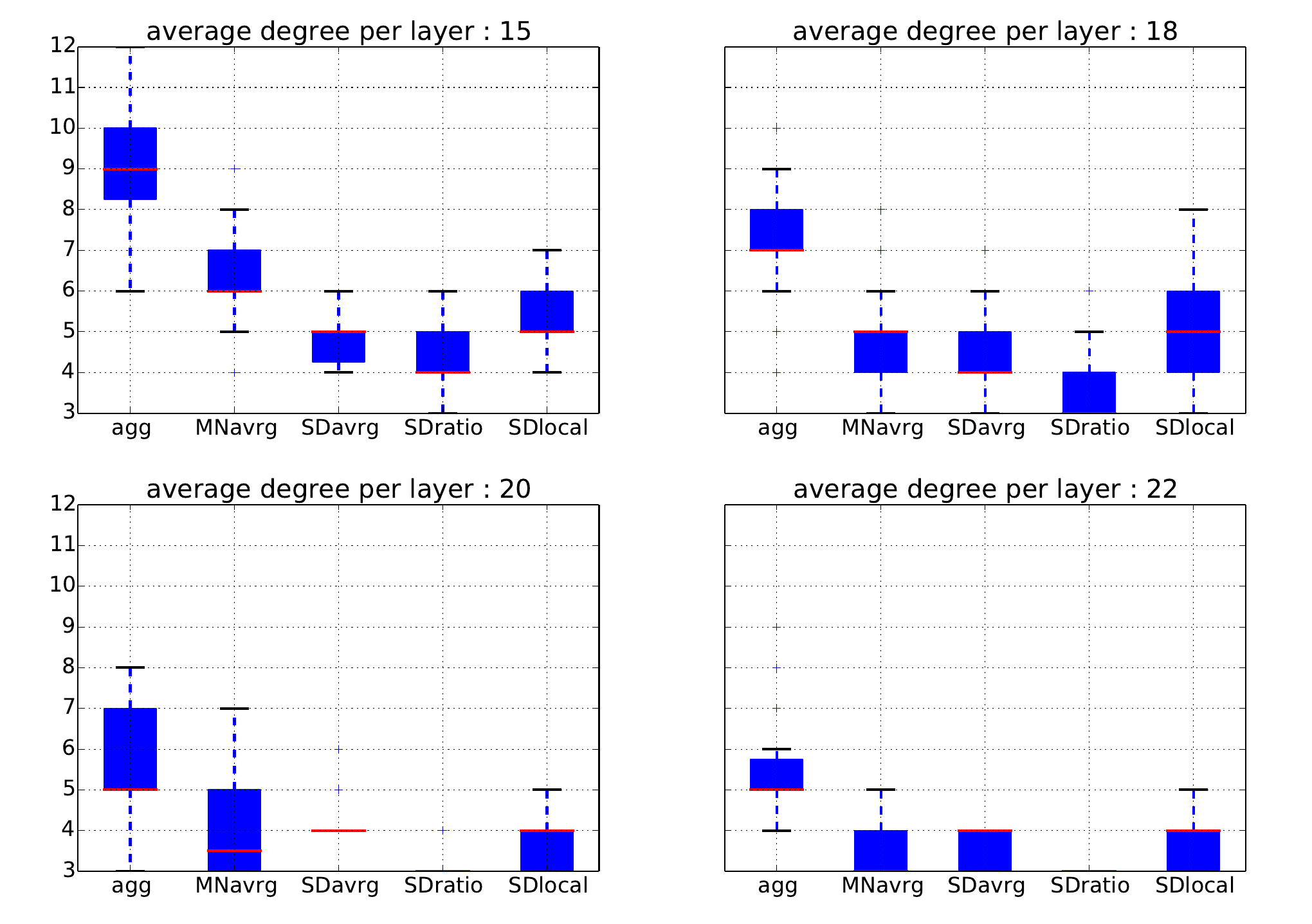}
\end{subfigure}
\vspace{-10pt}
\begin{center} (a) \end{center}
\vspace{-10pt}
\begin{subfigure}{0.48\textwidth}
\centering{}
\includegraphics[width=.95\linewidth]{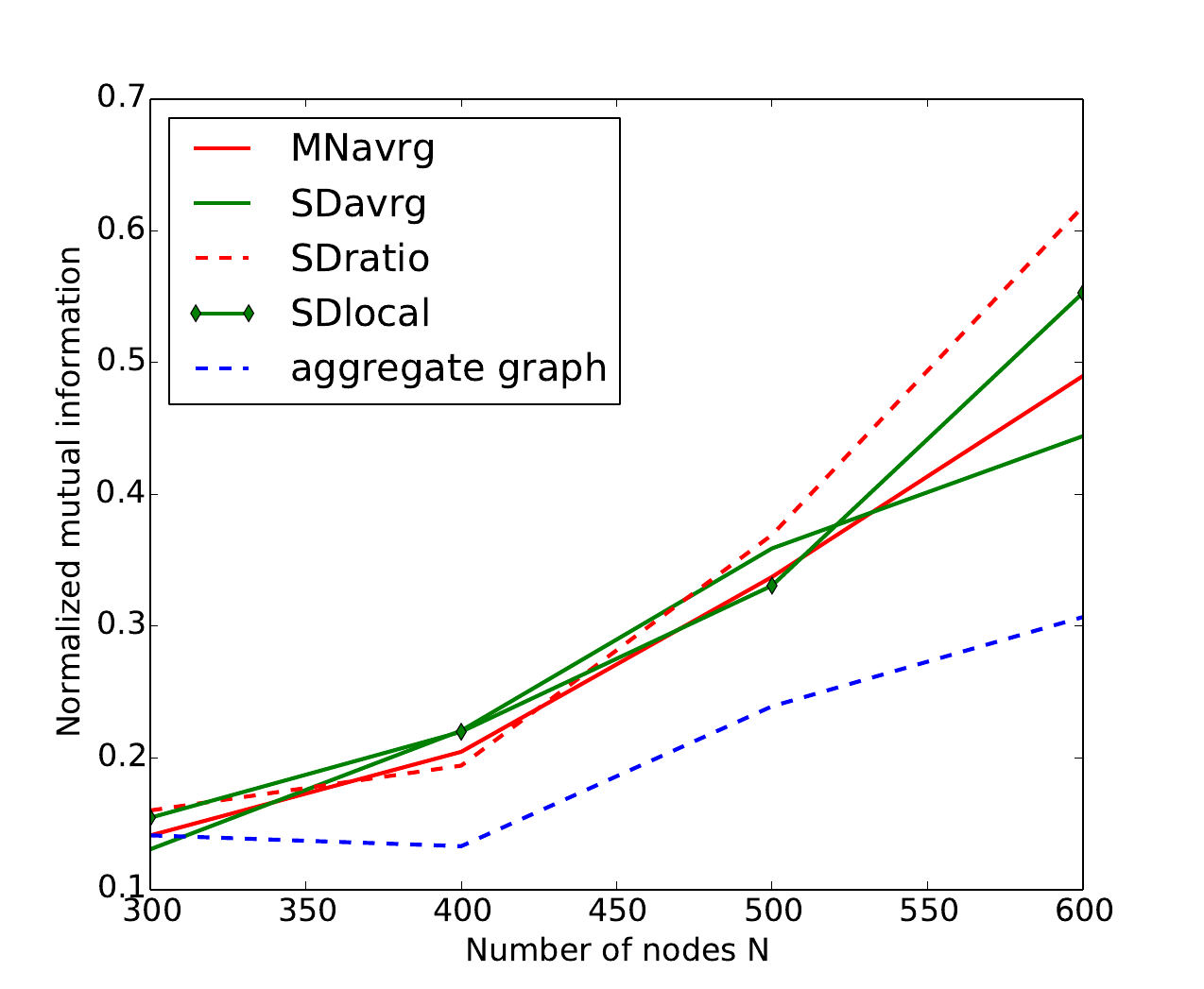}
\end{subfigure}%
\begin{subfigure}{0.48\textwidth}
\centering{}
\includegraphics[width=.95\linewidth]{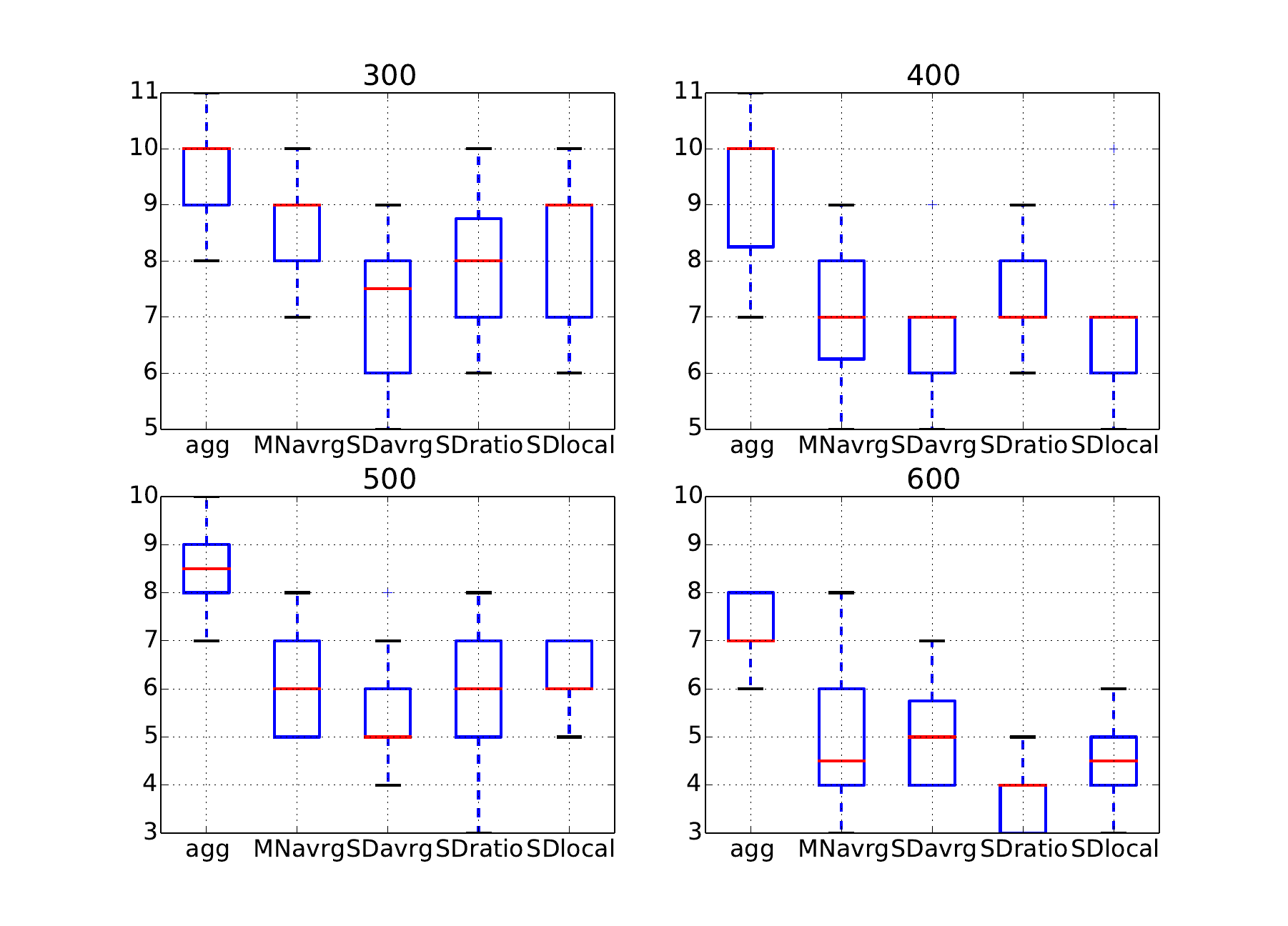}
\end{subfigure}
\vspace{-10pt}
\begin{center} (b) \end{center}
\vspace{-10pt}
\begin{subfigure}{0.48\textwidth}
\centering{}
\includegraphics[width=\linewidth]{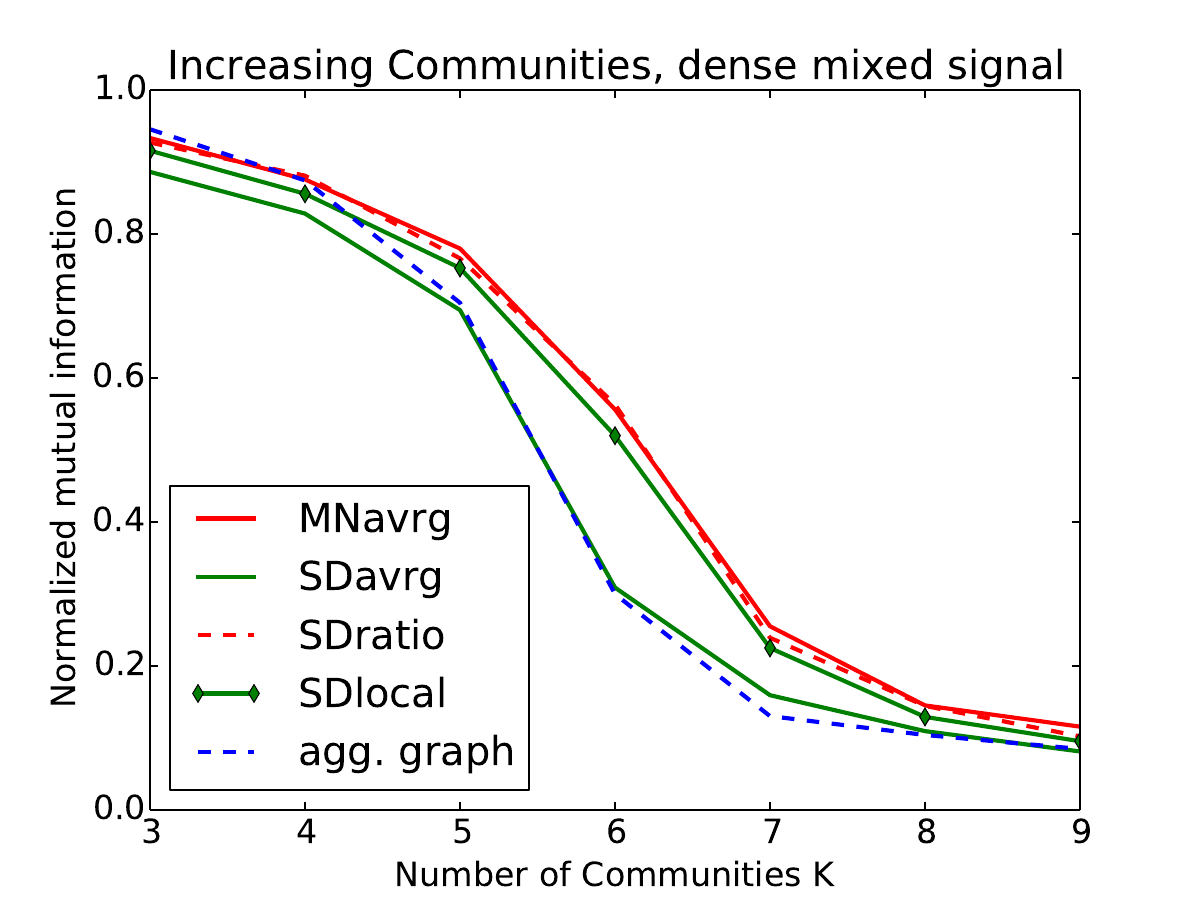}
\end{subfigure}%
\begin{subfigure}{0.48\textwidth}
\centering{}
\includegraphics[width=.95\linewidth]{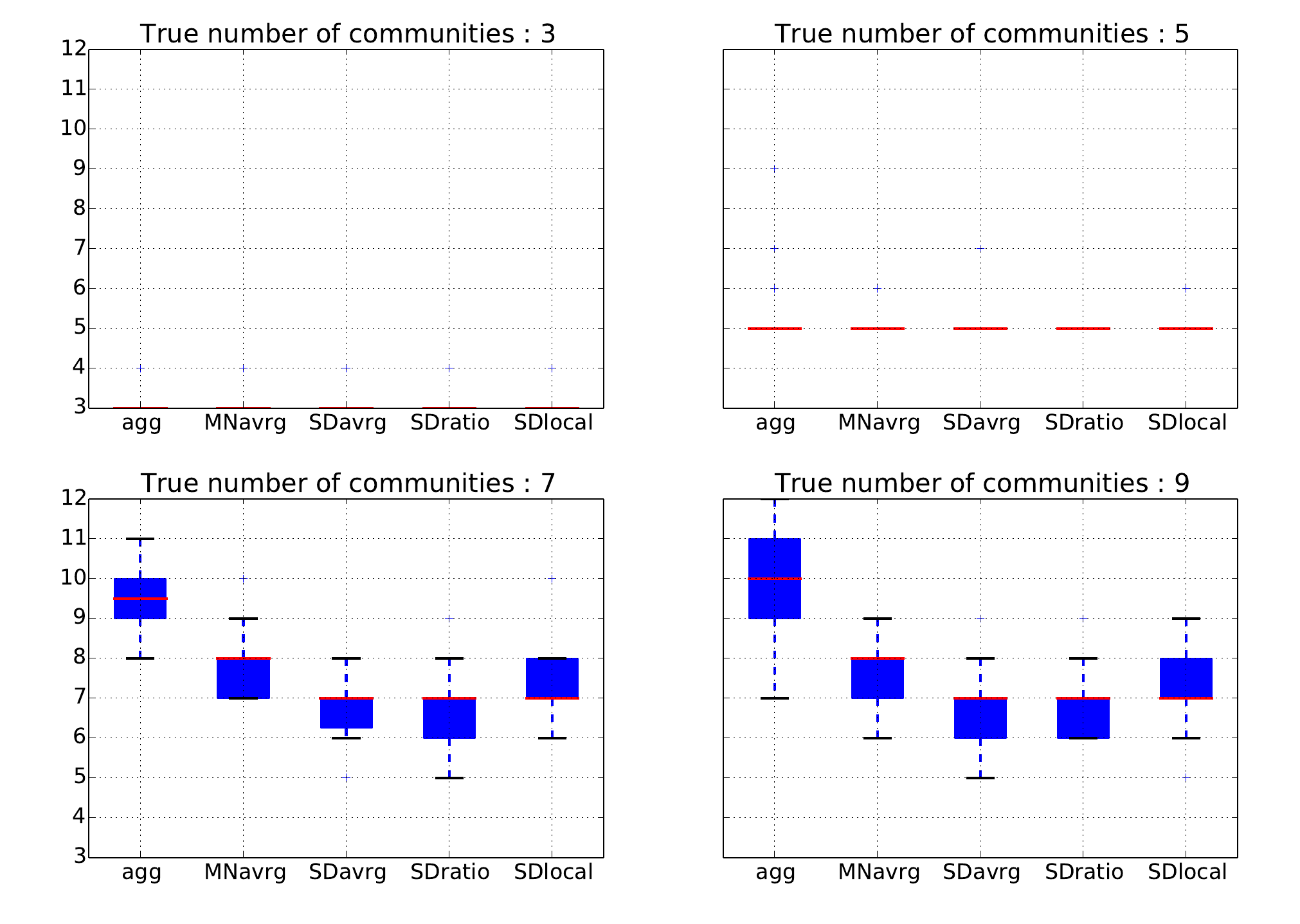}
\end{subfigure}
\vspace{-10pt}
\begin{center} (c) \end{center}
\vspace{-10pt}
\caption{Performance of various multi-layer modularities for data generated from the independent degree DCBM. Both sparsity and signal quality are mixed across different layers. (a) $N$ and $K$ are fixed at 800 and 3 while the average degree of the nodes across all layers combined is increased. (b) Increasing nodes with fixed $K=3$. (c) Increasing number of communities with fixed $N=800$.  In all cases, the left side figure is the comparison in terms of NMI and the right side figure is the box plot of the number of communities detected.}
\label{mixed}
\end{figure}

In this simulation, the component layers of the multi-layer graph vary in terms of both sparsity and signal strength in the following way: two layers are sparse and have strong signal, two layers are dense and have weak signal, while one layer is dense and have strong signal.  This scenario is extremely useful to test the methods in a very general situation where not all layers are informative and represent interactions that vary in scales. In real applications, we cannot expect all layers of a multi-layer network to contain high quality signals about the community structure. Some of the layers will have a high amount of noise interfering with the genuine signal from other layers.  Likewise the layers might represent very different relations and hence vary widely in density. We design three simulation setups: for the first one we vary the average density per layer while fixing $N$ and $K$ at 800 and 3 respectively,  for the second case we fix $K$ at 3 and let $N$ grow from 300 to 600 in steps of 100, and for the third one we fix $N$ at 800 and let the number of communities grow from 3 to 9. As with the previous case we report both the comparison in terms of NMI and box plots of the number of clusters detected.

For the first case, the results presented in Figure \ref{mixed}(a) show that MNavrg along with the shared degree methods outperform the aggregate graph consistently in terms of both NMI and the accuracy of the number of communities detected as the layers become denser. We observe a slight under-performance of MNarvg compared to SDlocal, SDratio and SDavrg in terms of the accuracy of the number of communities detection when the average density of layers is lower, but eventually their performance is comparable. Figure \ref{mixed}(b) shows very similar observation for the second case where the number of nodes is increasing. Finally Figure \ref{mixed}(c) shows that with increasing number of communities performance deteriorates in all the modularities, however the drop in performance is faster for aggregate graph and SDavrg compared to the others.

From this simulation we see that although the aggregate graph fails to provide good performance, the shared degree methods, in spite of combining information from all layers in their null model, performs at par with the MNavrg. Hence, this shows that the shared degree methods not only perform better in sparse networks, but are also robust against the presence of high degree of noise.

\subsection{Likelihood quality functions with number of communities known}

\begin{figure}[h]
\begin{subfigure}{0.5\textwidth}
\centering{}
\includegraphics[width=.95\linewidth]{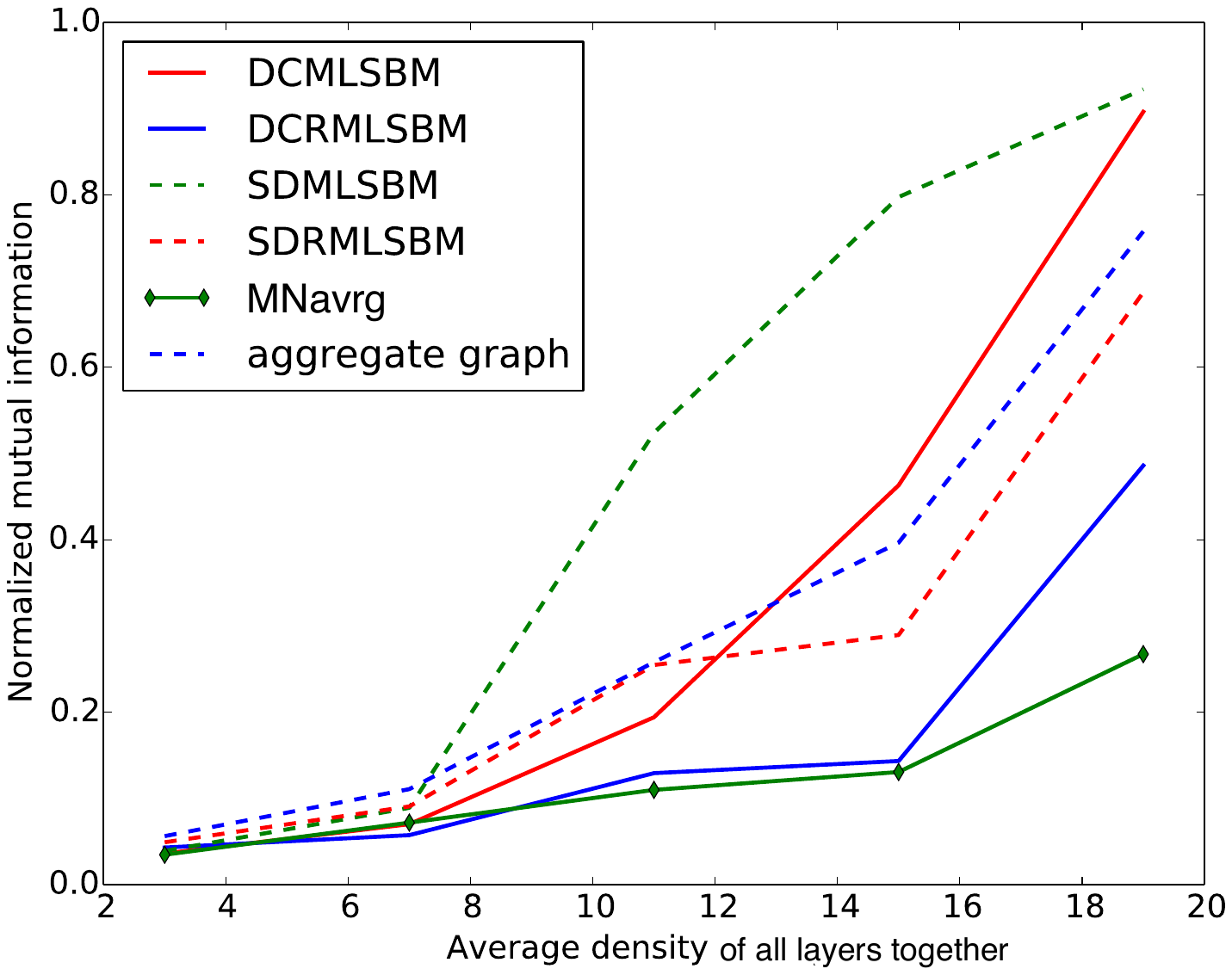}
\end{subfigure}%
\begin{subfigure}{0.5\textwidth}
\centering{}
\includegraphics[width=.95\linewidth]{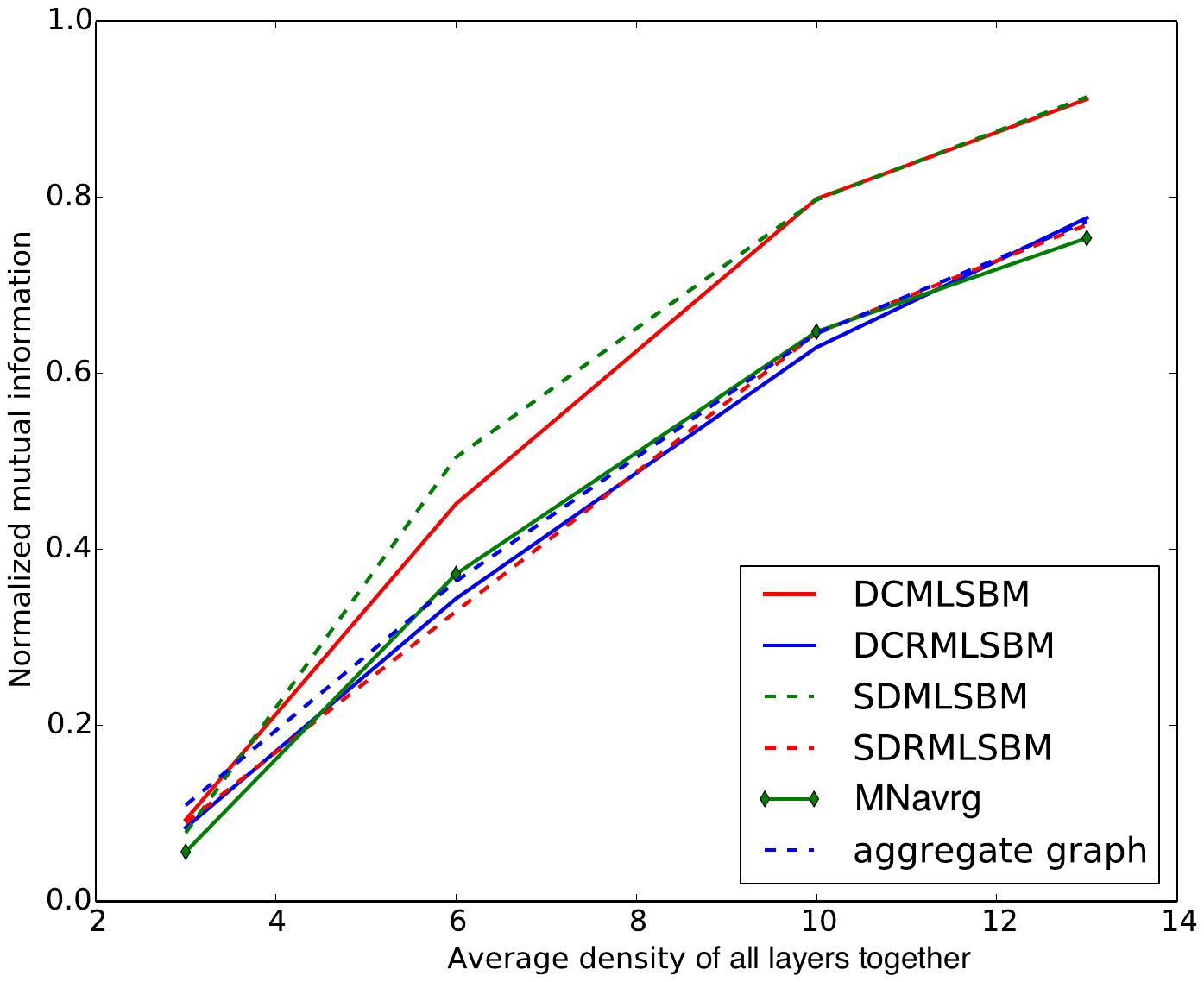}
\end{subfigure}
\vspace{-10pt}
\begin{center} (a) \hspace{200pt} (b) \end{center}
\vspace{-10pt}
\caption{Comparison of performance with known number of communities of various MLED modularities along with MNavrg for data generated from the  stochastic blockmodel. The layers are mixed in sparsity and signal quality. The average degree on nodes across all layers are increasing while $N$, $K$ and $M$ are fixed at 500, 2 and 4 respectively. We consider two cases: (a) balanced community sizes (roughly half of the nodes belonging to either cluster), and (b) unbalanced community sizes (30\% of the nodes belonging to one cluster and 70\% belonging to the other).}
\label{known}
\end{figure}

In this section we assume the number of communities $K$ to be known in advance and assess the effectiveness of the likelihood based quality functions. We also include two modularity functions, MNarvg and aggregate graph for comparison. For this simulation (Figure \ref{known}(a) and (b)) we fix $N$, $K$ and $M$ at 500, 2 and 4 respectively while we let the average degree density of all layers together to grow. In the first case (Figure \ref{known}(a)) the community sizes are balanced with roughly half of the nodes belonging to either cluster, and in the second case (Figure \ref{known}(b)) the community sizes are unbalanced with 30\% of the nodes belonging to one cluster and the remaining 70\% belonging to the other.  The layers are mixed in terms of density and signal quality. As mentioned in Section 5, the Kernighan-Lin type algorithm used when the number of communities is known requires an initial labeling. For this purpose we randomly permute the labels of 50\% of the nodes, keeping the correct labels for the rest of the nodes, similar to \citet{bc09}.

We notice that the DC-RMLSBM and SD-RMLSBM along with the aggregate graph method perform worse than the DC-MLSBM and SD-MLSBM as expected by the fact that the community signal content varies across layers. Moreover, as expected from our discussion in null model selection, we observe that the shared degree methods perform considerably better than the corresponding independent degree methods in this sparse regime. For example, we note that the shared degree method SD-MLSBM outperforms the corresponding independent degree method DC-MLSBM throughout the range of the simulation. The same can be observed comparing the performance of SD-RMLSBM with ID-RMLSBM, where SD-RMLSBM outperforms ID-RMLSBM for most of the range of the simulation. Lastly, the method MNavrg based on the multi-layer configuration model performs quite worse compared to the MLED likelihood quality function methods. This observation is consistent with the similar observations in case of single networks by \citet{bc09} and \citet{zlz12}.

\section{Real data analysis}

In this section we analyze a variety of multi-layer network datasets from different fields including social networks (three Twitter networks), friendship networks (Vickers-Chan's grade 7 peer network) and biological networks (Neuronal network of C-elegans). We demonstrate the effectiveness of the multi-layer methods discussed in this paper in detecting meaningful clusters in the networks.

\subsection{Twitter datasets}
We consider three real world multi-layer network datasets from the social network Twitter corresponding to interactions among (a) British Members of Parliaments (MPs), (b) Irish politicians and (c) Football players from the English premier league clubs. All the datasets were curated by \citet{gc13}. For each of the networks we consider three network layers corresponding to the twitter relations ``mentions", ``follows" and ``retweets" among a set of nodes. We apply the multi-layer community detection methods discussed in this paper to cluster the nodes. The ground truth community labels are also provided by \citet{gc13} and correspond to different underlying aspects of the nodes. For example, in the political networks (UK MPs and the Irish politicians) the ground truth corresponds to the political affiliation of the individuals, whereas in the network of premier league football players the ground truth corresponds to the teams (English premier league clubs) the players belong to.

\subsubsection*{UK MPs}
The first dataset consists of twitter interactions between 419 British MPs. We consider only those nodes which are connected by at least one connection in all the three layers. Then this is the intersection of the largest connected components in the layers. This reduces the number of nodes to 381. However, there are seven MPs in the trimmed network who do not belong to any major political party (named ``other" in the ground truth). Hence we remove those nodes and analyze the network with the remaining 374 nodes. The ground truth community assignment contains 152 Conservative, 178 Labour, 39 Liberal democrats and 5 SNP. The highly correlated layer-wise degree distribution of this network is presented in Figure \ref{fig:degree}. The number of communities detected and the NMI of the clustering result with ground truth for different community detection algorithms based on multi-layer configuration model modularities (optimized using Louvain type algorithm) are listed in Table \ref{tab:UK}. Clearly the multi-layer methods perform better than the single layer methods with several of the methods (MNavrg, SDlocal, NG modularity on aggregate graph) obtaining perfect clustering solution.

\begin{table}[!ht]
\caption{The number of communities detected and the NMI of clustering for different community detection methods for Twitter UK politics data. The community names are identified by optimal assignment.}
\begin{centering}
\begin{tabular}{ccccccc}
\hline
 Method & Conservative & Labour & Lib. Dem & SNP & no. comm. & NMI\tabularnewline
\hline
Ground truth & 152 & 178 & 39 & 5 & 4 & $-$\tabularnewline
Single (mention) & 149 & 172 & 48 & 5 & 4 & 0.8645\tabularnewline
Single (follow) & 152 & 177 & 45 & $-$  & 3 & 0.9644\tabularnewline
Single (retweet) & 153 & 173 & 41 & 7 & 4 & 0.8838\tabularnewline
MNavrg & 152 & 178 & 39 & 5 & 4 & 1.00\tabularnewline
SDarvg & 151 & 176 & 41 & 6 & 4 & 0.9601\tabularnewline
SDlocal & 152 & 178 & 39 & 5 & 4 & 1.00\tabularnewline
SDratio  & 152 & 178 & 44 & $-$& 3 & 0.9792\tabularnewline
Aggregate & 152 & 178 & 39 & 5 & 4 & 1.00\tabularnewline
\hline
\end{tabular}
\end{centering}
\label{tab:UK}
\end{table}

\subsubsection*{Irish politicians}
The dataset on Irish politicians consists of twitter interactions among 307 Irish politicians. The ground truth consists of party affiliations of them into Republic of Ireland's six major political parties. However 23 of them are independents and do not belong to any parties. We analyze the network both with and without these independents. The corresponding results are reported in Table \ref{tab:irish}(b) and \ref{tab:irish}(a) respectively. As expected the NMI with the ground truth is better when the network is analyzed without independents. In both cases, the multi-layer methods generally outperform the single layer methods. The highest NMI in both cases are obtained by the multi-layer method SDlocal. For the network without the independents, three multi-layer methods, MNavrg, SDlocal and SDratio, make only one incorrect assignment. 

When the number of communities is assumed to be known and set to 6, which is the number of communities in the ground truth, all multi-layer methods, based on both modularities and likelihood quality functions perform very well and miscluster only one node (Table \ref{tab:irish}(c)). Note in this case, with known $K$, the methods are optimized using Kernighan-Lin type algorithm.






\begin{table}[!ht]
\caption{The NMI of clustering from different community detection methods for Twitter Irish politics data. The community names are identified by optimal assignment, ``no. comm." stands for number of communities detected.}

\begin{centering}
\begin{tabular}{ccccccccc}
\hline
 Method & fg & Labour & ff & sf & ula & green & no. comm. & NMI  \tabularnewline
\hline
Ground truth & 124 & 69 & 45 & 31 & 8 &7  & 6 & $-$\tabularnewline
Single (mention) & 120 & 71 & 44 & 33 & 9 & 7 & 6 & 0.8901 \tabularnewline
Single (follow) & 124 & 72 & 49 & 39 & $-$& $-$ & 4 & 0.9353 \tabularnewline
Single (retweet) & 123 & 69 & 45 & 31 & 8 & 8 & 6 & 0.9763 \tabularnewline
MNavrg & 124 & 68 & 45 & 32 & 8 & 7 & 6 & \bf{0.9881} \tabularnewline
SDarvg & 125 & 67 & 45 & 32 & 8 & 7 & 6 & 0.9745 \tabularnewline
SDlocal & 124 & 68 & 45 & 32 & 8 & 7 & 6 & \bf{0.9881} \tabularnewline
SDratio & 124 & 68 & 45 & 32 & 8 & 7 & 6 & \bf{0.9881} \tabularnewline
Aggregate & 124 & 67 & 45 & 33 & 8 & 7 & 6 & 0.9796 \tabularnewline
\hline
\end{tabular}

\begin{center} (a) Without Independents \end{center}
\end{centering}

\smallskip
\begin{centering}
\begin{tabular}{cccccccccc}
\hline
 Method & fg & Labour & ff & sf & ula & green & ind &  no. comm. & NMI  \tabularnewline
\hline
Ground truth & 124 & 69 & 45 & 31 & 8 & 7 & 23  & 7 & $-$ \tabularnewline
Single (mention) & 116 & 70 & 47 & 38 & 9 & $-$& 27  & 6 &  0.8124 \tabularnewline
Single (follow) & 124 & 78 & 56 & 49 & $-$ & $-$ & $-$ & 4 & 0.8474\tabularnewline
single (retweet) & 122 & 79 & 46 & 32 & 8 & $-$& 20 & 6 & 0.8748\tabularnewline
MNavrg & 123 & 76 & 47 & 33 & $-$ & $-$& 28 & 5 & 0.8818  \tabularnewline
SDarvg  & 127 & 76 & 50 & 47 & 7 & $-$& $-$& 5 & 0.8502 \tabularnewline
SDlocal & 125 & 77 & 47 & 33 &  8 & $-$& 17 & 6 &  \bf{0.8927} \tabularnewline
SDratio & 125 & 76 & 50 & 47 & 9 & $-$& $-$& 5 & 0.8613 \tabularnewline
aggregate & 124 & 70 & 46 & 35 & 18 & $-$ & 14 & 6 & 0.8831
\tabularnewline
\hline
\end{tabular}
\begin{center} (b) With Independents \end{center}
\end{centering}

\begin{centering}
{\small
\begin{tabular}{ccccccc}
\hline
NG &  NG  & NG & DCBM  & DCBM  & DCBM  & aggregate \tabularnewline
(mention) & (follow) & (retweet) & (mention) & (follow) & (retweet) &
\tabularnewline
\hline
 0.9300 & 0.9353 & 0.9763 & 0.9289 & 0.9347 & 0.9763 & 0.9881\tabularnewline
 \hline
  \end{tabular}

 \medskip
 \begin{tabular}{cccccccc}
  \hline
DC-MLSBM &  DC-RMLSBM & SD-MLSBM & SD-RMLSBM & MNavrg & SDavrg & SDlocal & SDratio \tabularnewline
\hline
0.9881 &  0.9881 & 0.9881 & 0.9881 & 0.9881 & 0.9881 &0.9881 & 0.9881 \tabularnewline
\hline
\end{tabular}
}
\begin{center} (c) With known number of communities \end{center}
\end{centering}
 \label{tab:irish}
\end{table}

Our results on both political networks show how multi-layer methods can correctly identify meaningful community structure in networks. The near-optimal clustering for some of the multi-layer methods is quite surprising and quite rare in network community detection. This is perhaps an indication of how politicians heavily communicate with people within their political ideologies and seldom communicate with people of different ideologies. Hence the social interaction patterns of politicians easily reveal their political affiliations.

\subsubsection*{English premier league football players}

The last twitter dataset we analyze consists of interaction among sports personalities; the football players in  the English Premier League. As before, we keep only those nodes who are connected to at least one other node in each of the network layers. The ground truth for this dataset consist of footballers assigned to the 20 football clubs that they play for. The number of clusters detected along with the NMI of the solution with the ground truth are given in the Table \ref{tab:sports}(a). We see that almost all methods, single layer and multi-layer, underestimate the number of clusters. We compare the performance of these methods assuming that the number of clusters is known (20) in Table \ref{tab:sports}(b). We note that the MLCM based multi-layer methods clearly outperform not only the single layer modularities and the baseline aggregate method, but also the multi-layer block model modularities. The single layer NG modularities also outperform single layer DCBM modularities. In both groups of modularities, multi-layer modularities perform better than their single layer counterparts. Moreover one of the MLCM based shared degree method, the SDavrg, performs the best among all the methods. This is expected because when the number of communities $K$ is large, the number of parameters to be estimated in block model becomes large, resulting in poor estimation. Hence, the NG modularities outperform the block model ones, while the MLCM modularities outperform the multi-layer block model ones. For the same reason, among the multi-layer likelihood quality function methods, the restricted methods (DC-RMLSBM and SD-RMLSBM) with considerably less block model parameters to estimate perform better than the unrestricted ones.

\begin{table}[!ht]
\caption{Performance of different community detection methods in terms of (a) number of clusters detected and NMI of clustering, and (b) NMI of clustering with known number of communities for Twitter English Premier League dataset}

\begin{centering}
\begin{tabular}{ccc}
\hline
 Method & no. comm. & NMI \tabularnewline
\hline
Ground truth & 20 & $-$\tabularnewline
Single (mention) & 14 &  0.8104\tabularnewline
Single (follow) & 8 &  0.7656 \tabularnewline
Single (retweet) & 14 & 7550 \tabularnewline
MNavrg & 13 &  \bf{0.8330} \tabularnewline
SDarvg & 12 &  0.8105\tabularnewline
SDlocal & 12 &  0.8245\tabularnewline
SDratio  & 6 &  0.6996 \tabularnewline
Aggregate & 13 & 0.8204\tabularnewline
\hline
\end{tabular}
\quad  \quad
\begin{tabular}{cc}
\hline
 Method &  NMI \tabularnewline
\hline
NG (mention) &  0.8848 \tabularnewline
NG (follow) &  0.9022 \tabularnewline
NG (retweet) &  0.7910 \tabularnewline
DCBM (mention) &  0.7243 \tabularnewline
DCBM (follow) &  0.7552 \tabularnewline
DCBM (retweet) &  0.6765 \tabularnewline
DCMLSBM & 0.7898 \tabularnewline
DCRMLSBM & 0.8082 \tabularnewline
SDMLSBM & 0.7476 \tabularnewline
SDRMLSBM & 0.8125 \tabularnewline
MNavrg &  0.9176\tabularnewline
SDarvg &  \bf{0.9613}\tabularnewline
SDlocal & 0.9129\tabularnewline
SDratio & 0.9047\tabularnewline
Aggregate &  0.8896\tabularnewline
\hline
\end{tabular}

\end{centering}

\begin{center} (a) \hspace{200pt} (b) \end{center}
 \label{tab:sports}
\end{table}

\subsection{C-elegans}
Next we analyze a dataset from biology: the neuronal network connectome of a nematode Caenorhabditis elegans. It is the only organism whose wiring diagram or connectome of the entire nervous system is known and mapped \citep{chen06, white86}. For this dataset and the next one (grade 7 students) we use the versions of the dataset shared by \citet{muxviz14}. The present network consists of 279 neurons connected by two types of connections, a chemical link or synapse and an ionic channel \citep{nl14}, and is a weighted network. This network was previously analyzed both as a single layer network with the two layers collapsed together \citep{sohn11,varshney11,f07} and as a multi-layer network \citep{nl14}. We convert both layers into undirected network but keep the edge weights. Note that all our modularity measures can naturally handle positive edge weights with the weighted adjacency matrix replacing the binary adjacency matrix in all the calculations. Further, we consider only the nodes which are connected with at least one connection in both layers. The resulting network layers have 253 nodes and 1695 and 517 edges in the synapse and ion layers respectively.

\begin{figure}[!h]
\begin{subfigure}{0.5\textwidth}
\centering{}
\includegraphics[width=.95\linewidth]{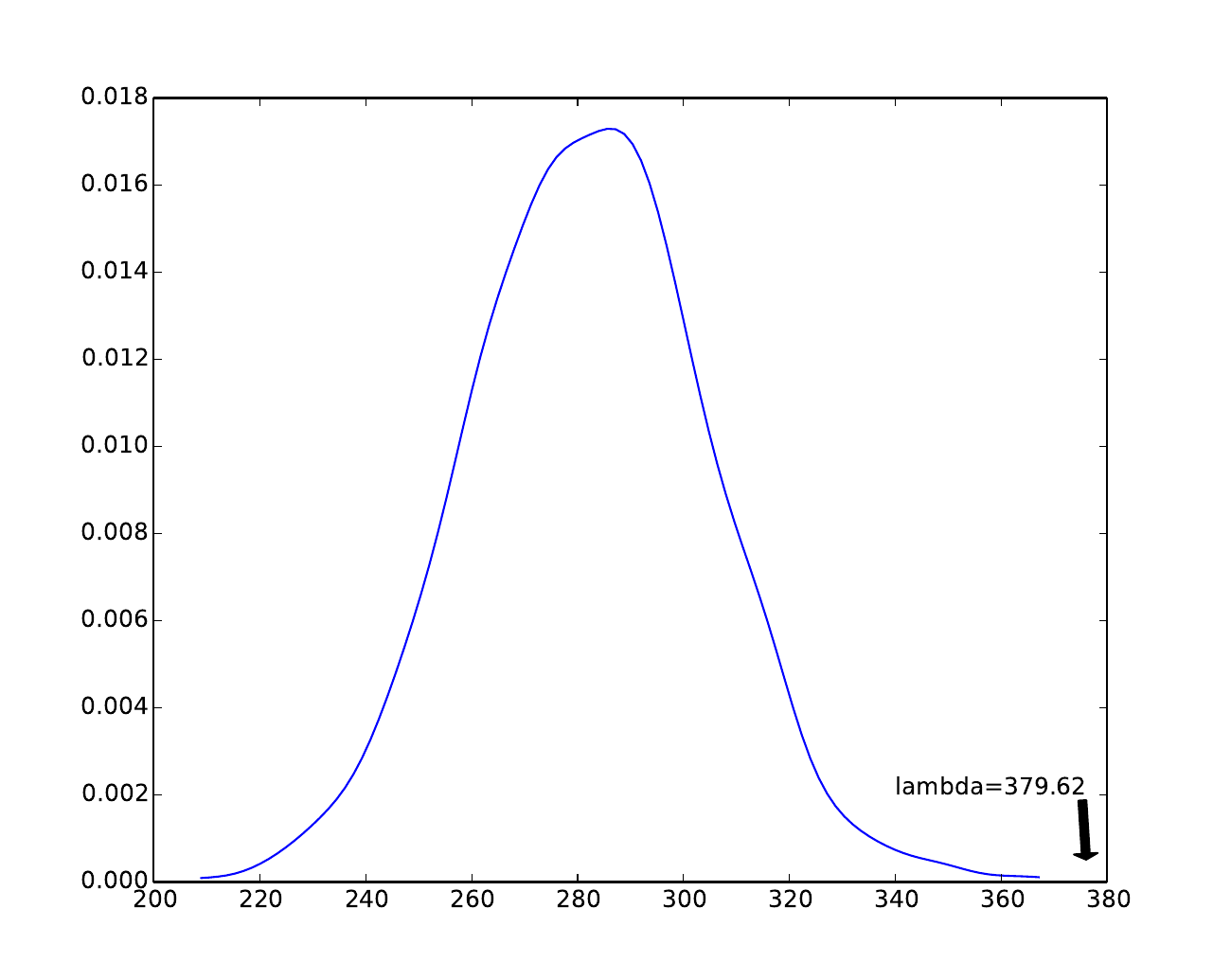}
\end{subfigure}%
\begin{subfigure}{0.5\textwidth}
\centering{}
\includegraphics[width=.95\linewidth]{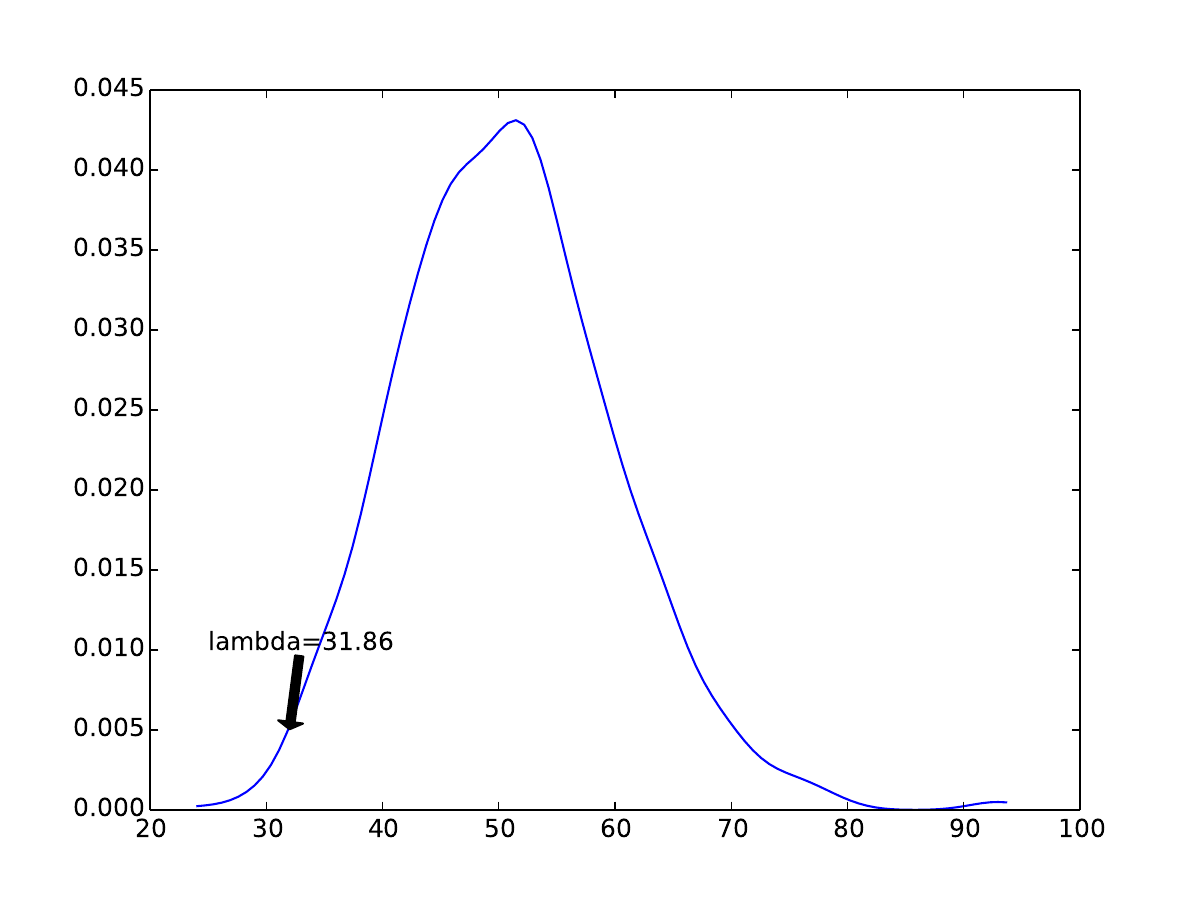}
\end{subfigure}
\begin{center} (a) \hspace{200pt} (b)  \end{center}
\caption{Parametric bootstrap distribution of the likelihood ratio test statistic for (a) C-elegans network and (b) Grade 7 students network. The observed value of the test statistic is indicated with an arrow.}
\label{bootstrap}
\end{figure}

\begin{figure}[h]
\centering{}

\includegraphics[width=0.6\linewidth]{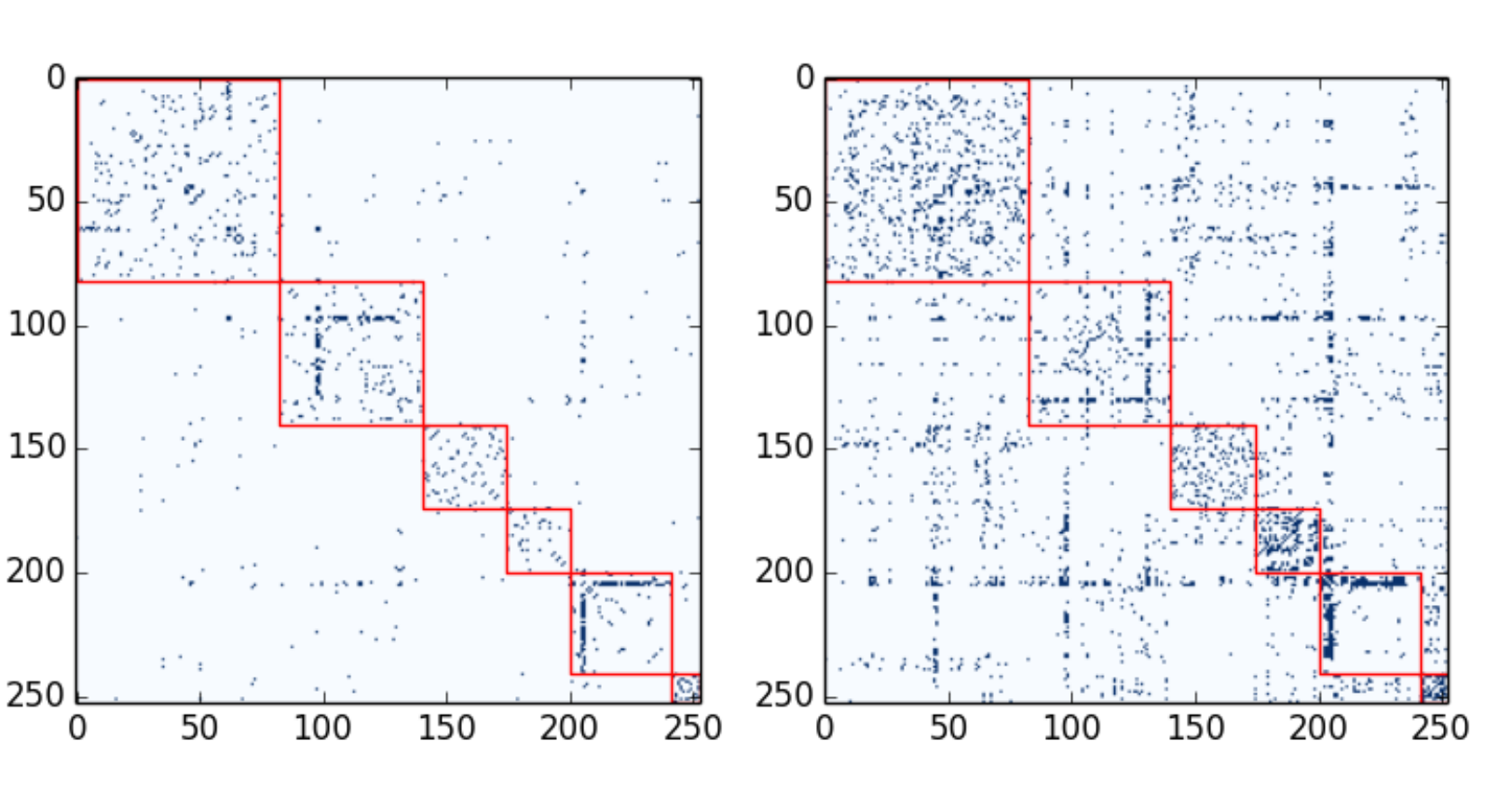}
\vspace{-10pt}
\begin{center} (a) \hspace{150pt} (b) \end{center}
\vspace{-10pt}
\caption{Adjacency matrices of the 2 layers in C-elegans connectome, (a) ionic channel and (b) chemical synapse, sorted and marked according to the clustering results obtained from SDlocal.} \label{fig:elegans_part5}
\end{figure}

\begin{figure}[h]
\centering{}
\includegraphics[width=0.6\linewidth]{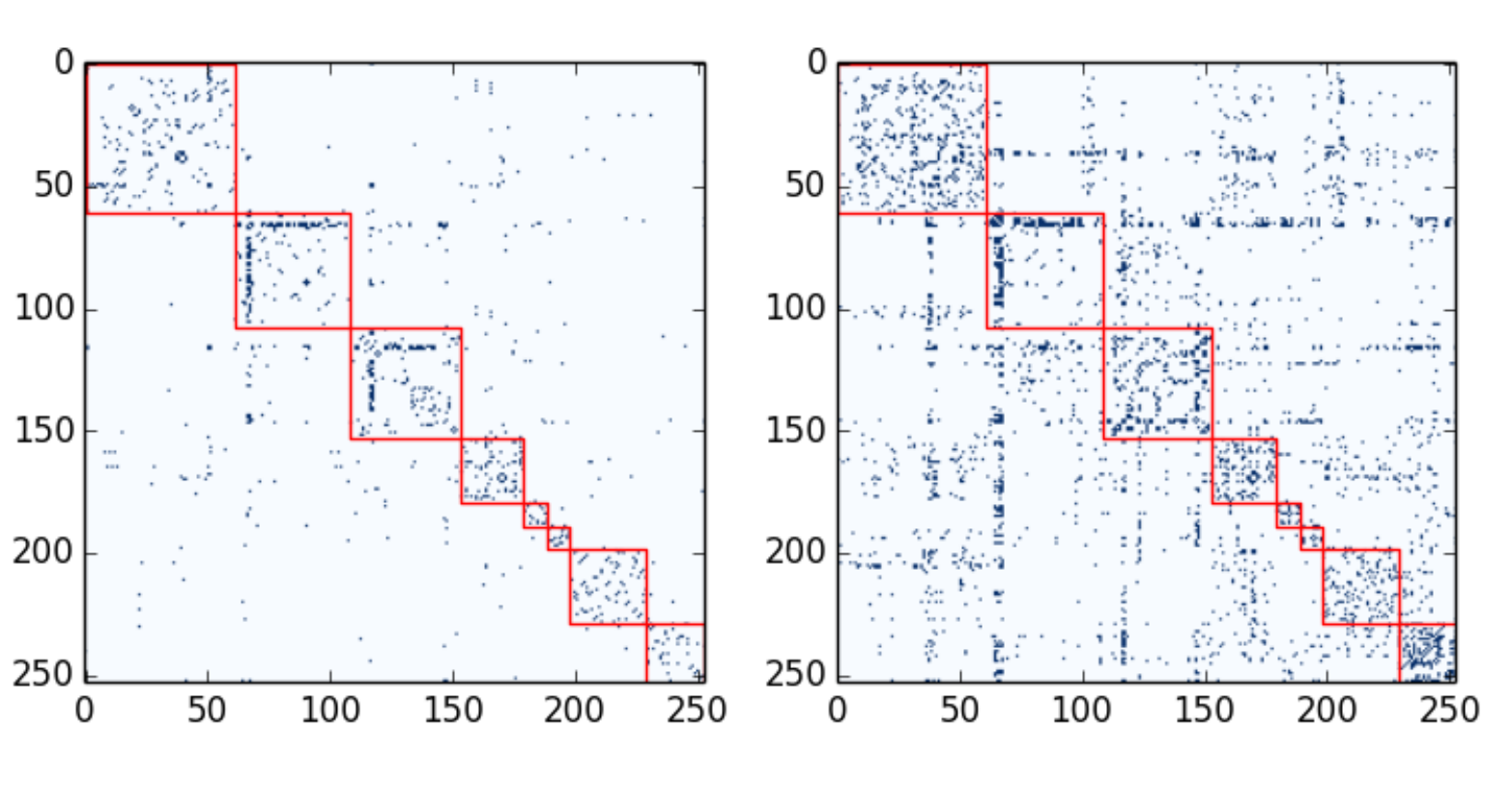}
\vspace{-10pt}
\begin{center} (a) \hspace{150pt} (b) \end{center}
\vspace{-10pt}
\caption{Adjacency matrices of the 2 layers in C-elegans connectome, (a) ionic channel and (b) chemical synapse, sorted and marked according to the clustering results obtained from MNavrg.} \label{fig:elegans_part1}
\end{figure}

We apply the hypothesis testing procedure developed in Section \ref{sec:modelselection} to test between SD and ID null models to this data. The LRT statistic value is 379.62. The parametric bootstrap distribution is shown in Figure \ref{bootstrap}(a). With an empirical $p$-value less than 0.01, we reject the null hypothesis of SD model and conclude that the ID model should be used as null model for community detection. Note that using the chi-squared distribution assumption (with degrees of freedom 252) for LRT statistic would also reach the same conclusion.
The two adjacency matrices plotted with class assignments from the multi-layer methods SDlocal and MNavrg  are presented in Figures \ref{fig:elegans_part5} and \ref{fig:elegans_part1} respectively. The block structure confirms well separated structural communities.

\subsection{Grade 7 students network}

This dataset, obtained by \citet{vc81}, is a multi-layer network on 29 students of grade 7. The students were asked to nominate a peer as an answer to one of the following three question: (a) Who do you get on with in the class? (b) Who are your best friends in the class? (c) Who would you prefer to work with? The answers to these three questions create three layers of relations among the students. Although the network edges are directed, we consider the network as a 3-layer undirected network and apply our community detection methods on it. The log-likelihood ratio test developed in Section \ref{sec:modelselection} fails to reject the null hypothesis of no difference between the shared degree and independent degree null models. The parametric bootstrap distribution is shown in Figure \ref{bootstrap}(b). The value of the LRT statistic is 31.86, which corresponds to a bootstrap $p$-value of 0.993. Moreover, the $p$-value obtained with a chi-squared approximation (with degrees of freedom 56) is 0.996, which is very close to the one obtained through bootstrap. Hence for parsimony we will prefer the shared degree null model.

\begin{figure}[!h]
\centering{}
\includegraphics[width=\linewidth]{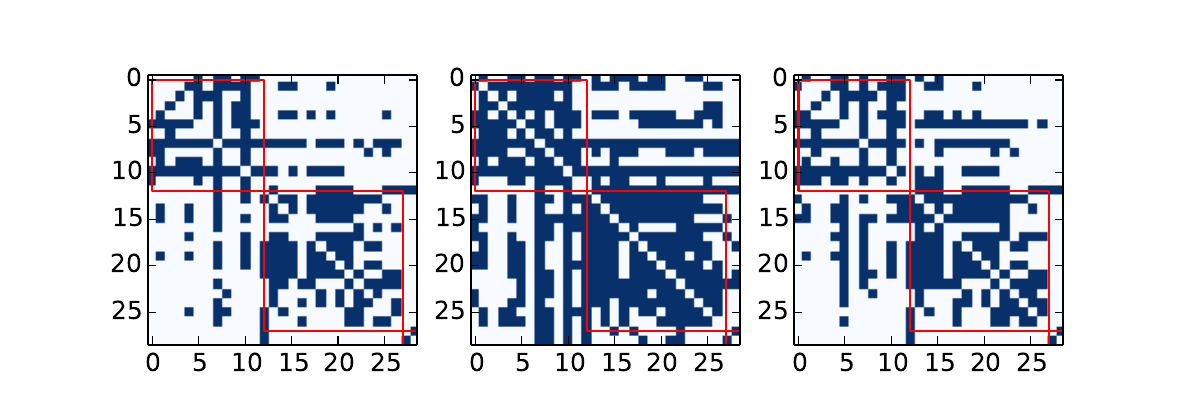}
\vspace{-20pt}
\begin{center} (a) \hspace{100pt} (b) \hspace{100pt} (c) \end{center}
\vspace{-10pt}
\caption{Adjacency matrices of the three layers, (a) get on with, (b) best friends and (c) work with, sorted and marked according to the (same) clustering result obtained from SDlocal, SDratio and MNavrg.} \label{fig:vickers}
\end{figure}

Single layer Newman-Grivan modularity gives 3, 4 and 3 clusters for get-on-with (gw), best friends (bf) and prefer to work with (ww) respectively. However using the entire multi-layer network, three of the four MLCM based methods along with aggregate detected 3 clusters of size 12, 15 and 2. Figure \ref{fig:vickers} depicts the three adjacency matrices sorted and marked into diagonal blocks according to this clustering solution. The density of intra-community edges are clearly higher than the inter-community edges across all three layers. Hence, the communities appear to be well separated in all three layers.

\begin{table}[!h]
\caption{NMI of clustering with the gender-wise clusters assumption for 7th grade students peer network; (a) Number of clusters detected and NMI of clustering  and (b) NMI of clustering with number of clusters given as 2.}

\begin{centering}
\begin{tabular}{ccc}
\hline
 Method & no. comm. & NMI \tabularnewline
\hline
Ground truth & 2 & $-$\tabularnewline
Single (gw) & 3 &  0.4698\tabularnewline
Single (bf) & 4 &  0.5871\tabularnewline
single (ww) & 3 & 0.5569\tabularnewline
MNavrg & 3 &  0.8726\tabularnewline
SDarvg & 2 &  0.7007\tabularnewline
SDlocal & 3 &  0.8726\tabularnewline
SDratio  & 3 &  0.8726\tabularnewline
aggregate & 3 & 0.8726\tabularnewline
\hline
\end{tabular}
\quad \quad
\begin{tabular}{cc}
\hline
 Method & NMI \tabularnewline
\hline
NG (gw) &  0.7007 \tabularnewline
NG (bf) &  1\tabularnewline
NG (ww) & 0.7007 \tabularnewline
DCBM (gw) &  0.4436 \tabularnewline
DCBM (bf) &  1\tabularnewline
DCBM (ww) & 0.8123 \tabularnewline
DCMLSBM & 1 \tabularnewline
DCRMLSBM & 1 \tabularnewline
SDMLSBM & 1 \tabularnewline
SDRMLSBM & 1 \tabularnewline
MNavrg & 1\tabularnewline
SDavrg & 0.8150 \tabularnewline
SDlocal & 1 \tabularnewline
aggregate & 1\tabularnewline
\hline
\end{tabular}
\end{centering}
\begin{center} (a) \hspace{200pt} (b) \end{center}
 \label{tab:vickersnmi}
\end{table}

Since the only external information known to us about these students is the gender information, we investigate how well the different clustering solutions align with communities based on genders. Surprisingly, we see quite high NMI for the clustering solution mentioned above (Table \ref{tab:vickersnmi}(a)). In fact, the nodes in the cluster of size 12 are all boys. The girls, however, got divided in to two classes, one of size 15 and another tiny cluster of size 2. In contrast, the three clustering solutions from the single layers yielded poor NMI with the gender-wise ground truth. From this we can conclude that fusing several layers of network information together, it is possible to learn meaningful information about the properties of the nodes, which would not have been possible with single layers. To further test our hypothesis of two gender-wise clusters in this multi-layer network, we employed the stochastic block model based modularities in conjunction with MLCM based modularities with known number of communities 2. The results (Table \ref{tab:vickersnmi}(b)) show all of the stochastic block model based modularities along with a number of MLCM based modularities perfectly agree with gender based ground truth.

\section{Discussions, Limitations, and Conclusions}
We have identified null models as the building blocks of modularity and likelihood quality function based community detection and introduced two sets of related multi-layer null models. The MLCM model conditions on observed degree vector sequence while the MLED model specifies an expected degree vector sequence. Both sets can be further divided into two categories, those based on independent degree principle and those based on shared degree principle. While the independent degree models have a separate degree parameter in each layer for each node, the shared degree null models ``share" the degree parameter across layers, with a layer specific parameter accounting for all heterogeneity. The shared degree null models have considerably fewer number of parameters and hence models based on them are more parsimonious. 
In this connection, we have also developed a hypothesis testing framework to test which model is more appropriate in a given scenario, an independent degree model or a shared degree model.

Several modularity and likelihood quality function measures have been derived based on these two sets of null models. Simulation results and real data applications show the effectiveness of these proposed methods in comparison to single layer methods and baseline procedures like applying single layer methods to an aggregate of the adjacency matrices of different layers. Based on our results, while we do not make any clear recommendation of a single measure to be used in all applications, we highlight some behaviors we observed and expect to observe under different situations. The shared degree models perform better in sparse graphs while the independent degree models perform better in relatively dense graphs. The likelihood quality function based methods generally perform either as good as or better than MLCM modularities and are suited for a more wider variety of networks. This is in line with the corresponding observation in single layer networks \citep{bc09,zlz12}.

When the number of communities are high and the layers are relatively sparse, the restricted block model based quality functions, DC-RMLSBM and SD-RMLSBM, perform better than the corresponding unrestricted ones, while the MLCM based modularities outperform both of these groups. This is because in those cases it is difficult to accurately estimate a large number of parameters that arise in block model based methods if $K$ is large. We also see that in such cases the shared degree versions of the models are more useful. The baseline aggregate of adjacency matrix although performs well under a few scenarios, it heavily relies on goodness of signal in denser layers and can not extract powerful signal from sparser layers. Hence aggregate works better mostly in situations where one or more of the comparatively denser layers also work well. 

\subsection*{Resolution limits}

The multi-layer modularities described in Section 3 further suffers from the well documented problem of resolution limit \citep{f07}. Briefly, the resolution limit refers to the property of modularity optimzation which prevents modules smaller in size than a limit from being detected even if the modular structure is strong. A common remedy is to use a resolution parameter. We will require similar resolution parameters for multi-layer modularity functions based on the multi-layer configuration model as well. However, it is not immediately clear if such an issue persists when likelihood quality functions from multi-layer degree corrected stochastic blockmodels are used. In fact it was shown in \cite{newman2016equivalence} that maximizing the likelihood of a restricted version of stochastic blockmodel (planted partition model) is equivalent to modularity optimization including the resolution parameter. 

\subsection*{Limitations}
Before we conclude, we point out a number of limitations for the approaches outlined in this paper.
First, the data structure we tackle in this manuscript is a subset of the more general multi-layer networks considered in parts of the literature that often contains entities of multiple types and inter-layer edges. We restricted our attention to the case when there are edges of different types among the same set of entities. Second, we focused ourselves on the problem of detecting a common community structure for the nodes taking into account the information contained in the whole multi-layer network. This is different from the problem of assigning separate but perhaps related communities to each node-layer pair.
Third, we acknowledge that our methods are suitable for networks generated only from the generative models we have proposed in this paper. The simulation studies we conducted assumed the networks to be generated from such models. Some of the models proposed elsewhere in the literature \citep{bazzi2020framework} are more general than our generative models, and hence our methods are not suitable for networks generated from those models. 

We conclude by pointing out that the principles of null models outlined in this paper can be extended to dynamic or time dependent networks as well and modularity scores can be developed based on suitable null models.

\section{Appendix}

\subsection{Proof of Theorem 1}

\begin{proof}
We start by noting that $\hat{\theta}_i = \frac{k_i}{\sqrt{2L}}$ and $\bar{\theta}_i = \frac{\kappa_i}{\sqrt{2\mathcal{L}}}$. From Chernoff inequality (Theorem A.1.4 of \cite{alon2004probabilistic}), we have for a given $i$,
\[
P(|k_i-\kappa_i| > \epsilon\sqrt{N\bar{\kappa}}) \leq 2\exp\left(-\frac{2\epsilon^2N\bar{\kappa}}{N}\right)=2\exp(-2\epsilon^2\bar{\kappa}).
\]
Taking a union bound over all $i$,
\[
P\left( \sup_{i \in \{1,\ldots,N\}} |k_i-\kappa_i| > \epsilon\sqrt{N\bar{\kappa}}\right) \leq 2N \exp(-2\epsilon^2\bar{\kappa})=\exp(\log (2N)-2\epsilon^2\bar{\kappa}) \to 0, \text{ as } N\to \infty,
\]
for a sufficiently large $C$ since $\bar{\kappa} \geq C \log N$ by assumption.
Therefore,
\begin{equation}
P\left(\sup_{i \in \{1,\ldots,N\}} \frac{|k_i-\kappa_i|}{\sqrt{2\mathcal{L}}} >\epsilon\right) \to 0 \text{ as } N \to \infty.
\label{kiconv}
\end{equation}

Now
\begin{align*}
    P\left(\left|\frac{k_i}{\sqrt{2L}} -\frac{\kappa_i}{\sqrt{2\mathcal{L}}}\right| > \epsilon\right) & =  P\left(\left|\frac{k_i}{\sqrt{2L}} -\frac{k_i}{\sqrt{2\mathcal{L}}} + \frac{k_i}{\sqrt{2\mathcal{L}}} -  \frac{\kappa_i}{\sqrt{2\mathcal{L}}}\right| > \epsilon\right)\\
    & \leq P\left(\left|\frac{k_i}{\sqrt{2L}} -\frac{k_i}{\sqrt{2\mathcal{L}}}\right| >\epsilon/2\right) + P\left( \left|\frac{k_i}{\sqrt{2\mathcal{L}}} -  \frac{\kappa_i}{\sqrt{2\mathcal{L}}}\right| > \epsilon/2\right)\\
    & \leq P\left(\frac{k_i}{\sqrt{2\mathcal{L}}}\left|\frac{\sqrt{2\mathcal{L}}}{\sqrt{2L}}-1\right| >\epsilon/2\right) + P\left(\frac{|k_i - \kappa_i| }{\sqrt{2\mathcal{L}}}>\epsilon/2\right).
\end{align*}
Note that $2\mathcal{L} = n \bar{\kappa}$.
Then for any $i$, $\frac{k_i}{\sqrt{2\mathcal{L}}} = \frac{\kappa_i}{\sqrt{2\mathcal{L}}}+o_p(1) \leq 1 + o_p(1) = O_p(1)$, since $\frac{\kappa_i}{\sqrt{2\mathcal{L}}} = \theta_i \leq 1$ by model assumption.
Moreover, since $L=\sum_{i,j} L_{ij}$ is the sum of $N^2$ independent random variables, 
\begin{align*}
    P(|2L-2\mathcal{L}| >\epsilon 2\mathcal{L}) \leq \exp\left(-\frac{2\epsilon^2 n^2 \bar{\kappa}^2}{n^2}\right) \to 0 \text{ as } N \to \infty.
\end{align*}
Therefore,
$
\frac{2L}{2\mathcal{L}} \overset{p}{\to} 1.
$
Since the function $\frac{1}{\sqrt{x}}$ is continuous at $x=1$, by continuous mapping theorem
\[
\frac{\sqrt{2\mathcal{L}}}{\sqrt{2L}} \overset{p}{\to} 1.
\]
Therefore, the quantity
\begin{equation}
\frac{k_i}{\sqrt{2\mathcal{L}}}|\frac{\sqrt{2\mathcal{L}}}{\sqrt{2L}}-1| = O_p(1) o_p(1) = o_p(1) \text{ for all } i.
\label{firstpart}
\end{equation}

Combining (\ref{firstpart}) and (\ref{kiconv}) we have the result
\begin{align*}
    P\left(\sup_{i \in \{1,\ldots,N\}}  \left|\frac{k_i}{\sqrt{2L}} -\frac{\kappa_i}{\sqrt{2\mathcal{L}}}\right| > \epsilon\right)
    & \leq P\left(\sup_{i \in \{1,\ldots,N\}} \frac{k_i}{\sqrt{2\mathcal{L}}}\left|\frac{\sqrt{2\mathcal{L}}}{\sqrt{2L}}-1\right| >\epsilon/2\right) + P\left(\sup_{i \in \{1,\ldots,N\}}  \frac{|k_i - \kappa_i| }{\sqrt{2\mathcal{L}}}>\epsilon/2\right) \\
    & \to 0.
\end{align*}

\end{proof}

\subsection{Proof of Theorem 2}
\begin{proof}
We follow the same proof technique as in Theorem 1. Recall
\begin{equation}
\hat{\theta}^{(m)}_{i} = \frac{k_{i}^{(m)}}{\sqrt{2L^{(m)}}},    \quad \text{ and } \bar{\theta}^{(m)}_{i} = \frac{\kappa_{i}^{(m)}}{\sqrt{2\mathcal{L}^{(m)}}},
\end{equation} 
Since $A_{ij}^{(m)}$ are independent binary random variables, from Chernoff inequality, we have for a given $i$ and $m$ and $\epsilon>0$,
\[
P\left(|k_i^{(m)}- \kappa_i^{(m)}| > \epsilon\sqrt{2\mathcal{L}^{(m)}}\right) \leq 2\exp\left(-\frac{2\epsilon^2 2\mathcal{L}^{(m)}}{N}\right) \leq 2\exp\left(-\frac{4\epsilon^2 2\mathcal{L}^{'}}{N}\right),
\]
where $\mathcal{L'} =\min_m \mathcal{L}^{(m)} \geq C N \log (MN)$ by assumption.
Taking a union bound over all $i$ and $m$,
\begin{equation}
P\left( \sup_{\substack{i \in \{1,\ldots,N\}, \\
    m \in \{1,\ldots,M\}}} \frac{|k_i^{(m)}- \kappa_i^{(m)}|}{{\sqrt{2\mathcal{L}^{(m)}}}} > \epsilon\right) \leq 2NM \exp\left(-\frac{4\epsilon^2 \mathcal{L}'}{N}\right)=\exp(\log (2MN)-4\epsilon^2C \log (MN)) \to 0,
    \label{kimconv}
\end{equation}
for a sufficiently large $C$.

Now similar to the arguments in the proof of Theorem 1, for any $\epsilon>0$ and given $i$ and $m$,
\begin{align*}
    P\left(\left|\frac{k_i^{(m)}}{\sqrt{2L^{(m)}}} -\frac{\kappa_i^{(m)}}{\sqrt{2\mathcal{L}^{(m)}}}\right| > \epsilon\right) & \leq P\left(\frac{k_i^{(m)}}{\sqrt{2\mathcal{L}^{(m)}}}\left|\frac{\sqrt{2\mathcal{L}^{(m)}}}{\sqrt{2L^{(m)}}}-1\right| >\epsilon/2\right) + P\left(\frac{|k_i^{(m)} - \kappa_i^{(m)}| }{\sqrt{2\mathcal{L}^{(m)}}}>\epsilon/2\right).
\end{align*}
Taking supremum over $i$ and $m$ we have
\begin{align*}
    P\left(\sup_{\substack{i \in \{1,\ldots,N\}, \\
    m \in \{1,\ldots,M\}}}\left|\frac{k_i^{(m)}}{\sqrt{2L^{(m)}}} -\frac{\kappa_i^{(m)}}{\sqrt{2\mathcal{L}^{(m)}}}\right| > \epsilon\right) & \leq P\left(\sup_{\substack{i \in \{1,\ldots,N\}, \\
    m \in \{1,\ldots,M\}}}\frac{k_i^{(m)}}{\sqrt{2\mathcal{L}^{(m)}}}\left|\frac{\sqrt{2\mathcal{L}^{(m)}}}{\sqrt{2L^{(m)}}}-1\right| >\epsilon/2\right) \\
    & \quad + P\left(\sup_{\substack{i \in \{1,\ldots,N\}, \\
    m \in \{1,\ldots,M\}}}\frac{|k_i^{(m)} - \kappa_i^{(m)}| }{\sqrt{2\mathcal{L}^{(m)}}}>\epsilon/2\right).
\end{align*}

From  (\ref{kimconv}), we have $\frac{k_i^{(m)}}{\sqrt{2\mathcal{L}^{(m)}}} = \frac{\kappa_i^{(m)}}{\sqrt{2\mathcal{L}^{(m)}}}+o_p(1) \leq O_p(1) + o_p(1) = O_p(1)$, since $\frac{\kappa_i}{\sqrt{2\mathcal{L}}} = \theta_i \leq 1$ by model assumption. Moreover, since the convergence in (\ref{kimconv}) holds for all $i$ and $m$, this result also holds for all $i$ and $m$.
Finally, 
\begin{align*}
    P\left( \sup_{m \in \{1,\ldots,M\}} \{|2L^{(m)}-2\mathcal{L}^{(m)}| >\epsilon 2\mathcal{L}^{(m)}\}\right) \leq \exp\left(\log M-\frac{8\epsilon^2 (\mathcal{L}^{'})^2}{N^2}\right) \to 0,
\end{align*}
since $\mathcal{L'} =\min_m \mathcal{L}^{(m)} \geq C N \log (MN).$
Therefore, 
\[
\sup_{m \in \{1,\ldots,M\}}\sqrt{\frac{2L^{(m)}}{2\mathcal{L}^{(m)}}} \overset{p}{\to} 1.
\]
Therefore,
\begin{align}
    \sup_{\substack{i \in \{1,\ldots,N\}, \\
    m \in \{1,\ldots,M\}}}\frac{k_i^{(m)}}{\sqrt{2\mathcal{L}^{(m)}}}\left|\frac{\sqrt{2\mathcal{L}^{(m)}}}{\sqrt{2L}^{(m)}}-1\right| = O_p(1)o_p(1) = o_p(1).
    \label{th2t1}
\end{align}
Hence combining results in (\ref{kimconv}) and (\ref{th2t1}) we have the desired result.

\end{proof}

\subsection{Proof of Theorem 3}

\begin{proof}
In the notation of the theorem, 
\begin{equation}
    \hat{\theta}_{i}=\frac{\sum_{m} k_{i}^{(m)}}{\sqrt{2L}}, \quad  \hat{\beta}_{m}=\frac{L^{(m)}}{L}, \quad \quad \text{ and } \quad \quad
        \bar{\theta}_{i}=\frac{\sum_{m} \kappa_{i}^{(m)}}{\sqrt{2\mathcal{L}}}, \quad  \bar{\beta}_{m}=\frac{\mathcal{L}^{(m)}}{\mathcal{L}}.
\end{equation}
 Note that by assumption, $\mathcal{L} = \sum_m \mathcal{L}^{(m)} \geq C NM \log N$. Since $A_{ij}^{(m)}$ are independent binary random variables, and $\sum_m k_i^{(m)} = \sum_m \sum_j A_{ij}^{(m)}$, from Chernoff inequality, we have for any $i$ and for any $\epsilon>0$,
\[
P\left(|\sum_m k_i^{(m)}- \sum_m \kappa_i^{(m)}| > \epsilon\sqrt{2\mathcal{L}}\right) \leq 2\exp\left(-\frac{2\epsilon^2 2\mathcal{L}}{NM}\right).
\]
Taking a union bound over all $i$,
\[
P\left( \sup_{i \in \{1,\ldots,N\}} |\sum_m k_i^{(m)}- \sum_m \kappa_i^{(m)}| > \epsilon\sqrt{2\mathcal{L}}\right) \leq 2N \exp\left(-\frac{4\epsilon^2 \mathcal{L}}{NM}\right) \leq \exp(\log(2N)-4\epsilon^2C \log N) \to 0,
\]
as $ N\to \infty$, for a sufficiently large $C$.
Therefore,
\[
P\left(\sup_{i \in \{1,\ldots,N\}} \frac{|\sum_m k_i^{(m)}- \sum_m \kappa_i^{(m)}|}{\sqrt{2\mathcal{L}}} >\epsilon\right) \to 0 \text{ as } N \to \infty.
\]

As a consequence of the above result, we have $\frac{\sum_m k_i^{(m)}}{\sqrt{2\mathcal{L}}} = \frac{\sum_m \kappa_i^{(m)}}{\sqrt{2\mathcal{L}}}+o_p(1) =O_p(1)$ (in particular, bounded by 2 with high probability) for all $i$, since $\frac{\sum_m \kappa_i^{(m)}}{\sqrt{2\mathcal{L}}} = \bar{\theta}_i \leq 1$ by model assumption.
Further,
\begin{align}
    P(|2L-2\mathcal{L}| >\epsilon 2\mathcal{L}) \leq 2 \exp\left(-\frac{2\epsilon^2 N^2M^2 (\log N)^2}{N^2M}\right) \to 0.
    \label{Lconv}
\end{align}

Then similar arguments as the proof of Theorem 1 lead to the result
\begin{align*}
    P\left(\sup_{i \in \{1,\ldots,N\}}  |\frac{\ \sum_m k_i^{(m)}}{\sqrt{2L}} -\frac{\sum_m \kappa_i^{(m)}}{\sqrt{2\mathcal{L}}}| > \epsilon\right) \to 0.
\end{align*}

Next we prove the result for the estimators of the $\beta_m$ parameters. Clearly, since $L^{(m)}$ is the sum of $N^2$ independent random variables,
\begin{align}
    P\left( \sup_{m \in \{1,\ldots,M\}} \{|2L^{(m)}-2\mathcal{L}^{(m)}| >\epsilon 2\mathcal{L}\}\right) \leq \exp\left(\log M-\frac{8\epsilon^2 \mathcal{L}^2}{N^2}\right) \to 0,
    \label{Lmconv}
\end{align}
since $\mathcal{L} \geq C NM \log N.$

On the other hand, (\ref{Lconv}) shows that 
\[
\left|\frac{L}{\mathcal{L}}-1\right| =o_p(1).
\]
Now
\begin{align}
    P\left(\left|\frac{L^{(m)}}{L} -\frac{\mathcal{L}^{(m)}}{\mathcal{L}}\right| > \epsilon\right) 
    & \leq P\left(\left|\frac{L^{(m)}- \mathcal{L}^{(m)}}{\mathcal{L}}\right| >\epsilon/2\right) + P\left( \frac{L^{(m)}}{\mathcal{L}}\left|\frac{\mathcal{L}}{L}-1\right| >\epsilon/2\right).
    \label{th3final}
\end{align}
Since for any $m$, $ L^{(m)} =\mathcal{L}^{(m)} + o_p(1)$, and $\mathcal{L}^{(m)} \leq \mathcal{L}$, we have $\frac{L^{(m)}}{\mathcal{L}} =O_p(1)$, i.e., bounded (by 2) in high probability. Therefore, in the last term of (\ref{th3final}),
\[
\frac{L^{(m)}}{\mathcal{L}}\left|\frac{\mathcal{L}}{L}-1\right| =O_p(1)o_p(1) = o_p(1),
\]
for any $m$, while in first term on the right hand side of (\ref{th3final}), $\left|\frac{L^{(m)}- \mathcal{L}^{(m)}}{\mathcal{L}}\right|$ is also $o_p(1)$ for any $m$ by (\ref{Lmconv}). Therefore, combining the two results leads to the result.
\end{proof}

\subsection{Approximations without assuming self-loops}
While the model with self-loops is commonly used in the literature due to simplified computations \citep{arcolano2012moments,kn11,newman2016equivalence}, we do note that such a model may not be appropriate for graphs that do not contain self-loops. Here we estimate the expected error in the estimators if the model does not allow for self-loops.
For the ID model, plugging in the proposed estimator into the likelihood equations leads to 
\[
 \frac{\sum_j A_{ij}^{(m)}}{\hat{\theta}^{(m)}_i} - \sum_{j} \hat{\theta}_j^{(m)} + \hat{\theta}_i^{(m)}= \frac{k_{i}^{(m)}}{\sqrt{2L^{(m)}}}.
\]
The expected error can be approximated with standard assumptions on growth rates of degrees widely employed in the literature. First we note that a first order Taylor series approximation gives
\[
E[\hat{\theta}^{(m)}_{i}] = E\left[\frac{k_{i}^{(m)}}{\sqrt{2L^{(m)}}}\right] \approx \frac{E[k_{i}^{(m)}]}{\sqrt{2E[L^{(m)}}]}.
\]
It is common in the literature to assume that expected degrees in sparse networks scale with $O(\log N)$. Therefore $E[k_i^{(m)}] = O(\log N)$ and $E[L^{(m)}] = O( N\log N)$. Therefore the extent of error in each of the likelihood equation is $O(\sqrt{\frac{\log N}{N}})$.

Plugging in the estimators for the SD model in the likelihood equation for the SD model leads to the following estimate of errors:
\begin{align*}
        \frac{\partial l }{\partial \theta_i} & :\ \  \frac{\sum_m \sum_j A_{ij}^{(m)}}{\hat{\theta}_i} - \sum_m\sum_{j} \hat{\theta}_j + \sum_m \hat{\theta}_i = \sum_m \hat{\theta}_i, \quad \quad i = \{1,\ldots N\}, \\
    \frac{\partial l }{\partial \beta_m} & :\ \  \frac{\sum_{i<j} A_{ij}^{(m)}}{\hat{\beta}_m} - \sum_{i<j } \hat{\theta}_i\hat{\theta}_j = L-L = 0, \quad \quad  m=\{1,\ldots, M\}.
\end{align*}
Therefore there is no error in the second set of likelihood equations and the error in the first set can be quantified with the above growth rate assumptions. In particular the extent of error in each of the likelihood equations in the first set is  $O(M)$.

\bibliography{modularity}

\end{document}